\documentstyle[epsf,aps,psfig]{revtex}
\renewcommand{\thefootnote}{\fnsymbol{footnote}}
\begin{document}
\vspace*{1cm} 
\setcounter{footnote}{1}
\begin{center}
{\Large\bf Superfluidity in a Model of Massless Fermions
Coupled to Scalar Bosons}
\\[1cm]
Robert D.\ Pisarski $^1$ and Dirk H.\ Rischke $^2$
\\ ~~ \\
{\it $^1$Department of Physics} \\
{\it Brookhaven National Laboratory, Upton, New York 11973, U.S.A.} \\
{\it email: pisarski@bnl.gov} 
\\ ~~ \\
{\it $^2$ RIKEN-BNL Research Center} \\
{\it Brookhaven National Laboratory, Upton, New York 11973, U.S.A.}\\
{\it email: rischke@bnl.gov} 
\\ ~~ \\ ~~ \\
\end{center}
\begin{abstract} 
We study superfluidity in a model of massless fermions coupled 
to a massive scalar field through a Yukawa interaction.
Gap equations for a condensate with total spin $J = 0$ are solved
in the mean-field approximation.
For the Yukawa interaction, the gaps for right- and left-handed
fermions are equal in magnitude and opposite in sign, so that
condensation occurs in the $J^P =0^+$ channel.
At finite scalar mass, there are two different gaps for fermions
of a given chirality, corresponding to condensation of particle pairs or
of antiparticle pairs. These gaps become degenerate
in the limit of infinite scalar mass.
\end{abstract}
\renewcommand{\thefootnote}{\arabic{footnote}}
\setcounter{footnote}{0}

\section{Introduction}

In fermionic matter which is sufficiently cold and dense, 
any attractive interaction at the Fermi surface leads to the formation 
of Cooper pairs \cite{fetterwalecka}. The Cooper pairs form a Bose 
condensate, so that exciting a quasiparticle costs
an amount of energy $\geq 2|\phi|$, where $|\phi|$ is the gap energy.
This gap produces superfluidity, or, if the fermions are coupled
to a gauge field, superconductivity.

In quantum chromodynamics (QCD), one-gluon exchange between two
quarks is attractive in the color-antitriplet channel. 
One therefore expects that at sufficiently large quark chemical 
potential $\mu$ and sufficiently small temperature $T$ quarks 
condense into Cooper pairs which are color antitriplets.
This condensate breaks the $SU(3)$ color symmetry of the ground state, 
and gives rise to {\em color superconductivity}. 
For QCD, this phenomenon was first investigated by Barrois \cite{barrois},  
by Bailin and Love \cite{bailinlove}, and others \cite{others}. 
In seminal work, Bailin and Love estimated 
the gap energy to be on the order of $|\phi| \sim 10^{-3} \mu$.
Since the critical temperature for the onset of superconductivity,
$T_c$, is of the order of the gap energy, $|\phi|$, 
a color-superconducting phase of quark matter
could form in the interior of neutron stars. 
Color superconductivity was also studied in an $SU(2)$ gauge theory
\cite{kond}.

Interest in this subject was renewed through works by Alford, 
Rajagopal, and Wilczek
\cite{arw} and by Rapp, Sch\"afer, Shuryak, and Velkovsky \cite{rssv}.
In these papers, the attractive interaction was modeled with the
instanton vertex.
In the framework of a simple model of the type studied by
Nambu and Jona-Lasinio (NJL) \cite{nambu}, these authors found gap energies
of the order of $|\phi| \sim \mu$. Since $T_c \sim |\phi|$, 
for gap energies of this order of magnitude
color superconductivity is no longer only relevant for astrophysical
scenarios, it could also occur in relativistic
heavy-ion collisions, whenever the bombarding energy is large enough
to form relatively cold, baryon-rich quark-gluon matter.
The maximum amount of baryon stopping was found in heavy-ion collisions
at AGS energies, $E_{\rm Lab} \sim 10$ AGeV; therefore, the most favorable
conditions to form a color-superconducting phase of quark-gluon matter
would occur in nuclear collisions somewhere in the energy range from 
$E_{\rm Lab} \sim 1$ to $40$ AGeV, i.e., from GSI--SIS to CERN--SPS energies.

Most studies of color superconductivity assumed that only the
two lightest quark flavors, $u$ and $d$, form Cooper pairs.
This is based on the argument that, 
due to the larger strange quark mass, $m_s \simeq 100$ MeV,
the Fermi surfaces of non-strange and strange quarks do not match 
\cite{bailinlove}, consequently there are no $us$ or $ds$ Cooper pairs. 
However, for a system where $\mu_s = \mu_u = \mu_d \equiv \mu$,
this mismatch is, to leading order in $\mu$, only of order 
$(k_{F,q} - k_{F,s})/\mu
\sim (m_s^2 - m_q^2)/2\mu^2$, and therefore vanishes at sufficiently
large chemical potential. 
Based on this observation, Alford, Rajagopal, and Wilczek
suggested a new form of color superconductivity, where all three
flavors condense to lock the breaking of color and flavor symmetry 
\cite{arw2}. 
Comparison with earlier investigations \cite{paterson} of the patterns
of symmetry breaking in chiral models revealed \cite{pr} 
that the color-flavor locked state is energetically favored.
This was independently confirmed by
Sch\"afer and Wilczek in a numerical study of the effective
potential \cite{sw2}.

Other studies which investigated
the interplay of color superconductivity and
chiral symmetry breaking at finite temperature include those of
Berges and Rajagopal \cite{br} and Langfeld and Rho \cite{lr}.
Color superconductivity was also investigated via
renormalization group techniques by Evans, Hsu, and Schwetz \cite{ehs}
and by Sch\"afer and Wilczek \cite{sw1}. 

All of these studies were based exclusively on NJL-type models.
The sole exception is the work of Son \cite{son} who applied 
renormalization group techniques
to study the scale of the energy gap in QCD.
He pointed out that non-instantaneous one-gluon exchange modifies
the weak-coupling expression for the gap, $|\phi| \sim \mu\, \exp(-c/g^2)$
to $|\phi| \sim \mu\, \exp(-c/g)$; this was also noted in \cite{pr}.
This increases the likelihood to find color superconductivity at nonzero
temperature.

A quark-quark condensate $\Delta^{ij}_{fg,\alpha \beta}$ 
is a $N_c \times N_c$ matrix in color space ($i,j  = 1,\ldots , N_c$),
a $N_f \times N_f$ matrix in flavor space ($f,g = 1,\ldots, N_f$),
and a $4 \times 4$ matrix in Dirac space ($\alpha, \beta = 1,\ldots, 4$).
Quark pair condensation can in principle occur in channels with
arbitrary total spin $J$. Most previous studies indicate, however,
that condensation in the channel with total spin $J=0$ is favored
for two or more flavors.

In this work, we shall not discuss the color or flavor structure
of the gap matrix \cite{QCDrdpdhr}, but rather focus on the Dirac structure. 
To do so, it is sufficient to study a simpler model, where
fermions interact via the exchange of scalar bosons,
\begin{equation} \label{L}
{\cal L} = \bar{\psi} \left( i \gamma \cdot \partial - m \right) \psi
- g \bar{\psi} \psi \, \phi + \frac{1}{2} \left( \partial_\mu \phi \,
\partial^{\mu} \phi - M_s^2 \phi^2 \right)\,\, .
\end{equation}
(We do not add a quartic self-interaction for the scalar field;
such an interaction will not qualitatively change our results.)
Scalar one-boson exchange is attractive, therefore, we expect the
formation of Cooper pairs at finite density and sufficiently low
temperature in the $J=0$ channel.
Since the fermions in the Lagrangian (\ref{L}) are not charged,
this model can only exhibit superfluidity, not superconductivity.
Nevertheless, the Dirac structure of the superfluid condensate
is closely analogous to that 
of the color-superconducting condensate in QCD. 

Our principal result is the following. Massless fermions have
four types of $J=0$ condensates, corresponding to the pairing of
fermions with the same helicity and chirality \cite{pr}. For fermions
interacting via Yukawa interactions as in (\ref{L}), 
the gap for right-handed, positive helicity quasiparticles is equal in
magnitude (and opposite in sign) to that for left-handed, negative 
helicity quasiparticles. A novel feature of the relativistic treatment is
the appearance of gaps for quasi-antiparticles. With Yukawa interactions,
we find that the gap for right-handed, negative helicity 
quasi-antiparticles is equal in magnitude to that for left-handed, 
positive helicity quasi-antiparticles. 
These relations between the right- and left-handed gaps imply that there 
is condensation only in the $J^P=0^+$ channel.
For $M_s < \infty$, the quasiparticle and 
quasi-antiparticle gaps are not equal.

In the limit $M_s \rightarrow \infty$, the model (\ref{L})
reduces to an NJL-type model, and the quasiparticle and
quasi-antiparticle gaps become degenerate. Thus NJL-type models
are unrealistic, in that they force the equality of the quasiparticle and
quasi-antiparticle gaps.

The number of condensates is not merely a technical matter, but can even 
affect the order of a superconducting phase transition. 
Consider the transition where up and down quarks condense to
a color-superconducting phase. 
In \cite{pr} we demonstrated that if only one condensate appears, such as that
for particles, the transition is in the universality class 
of a single $U(3)$ vector and can be of second order.
Based on the analysis of this paper, it seems unavoidable that
condensates for both particles and antiparticles
appear together. The universality class becomes that of two
$U(3)$ vectors; while the fixed-point structure of this model is unknown,
it might be driven first order by the Coleman--Weinberg phenomenon.
If true, it implies that for two degenerate flavors, the color-superconducting 
transition is of first order, independent of the chiral
transition \cite{br}.

In \cite{pr} we also showed that the transition where up and down
quarks condense with strange quarks is driven first order by the
Coleman--Weinberg mechanism.
Sch\"afer and Wilczek argued that the degrees of freedom
in quark and hadron matter match at nonzero $\mu$ \cite{sw2}.
Their observation implies that this line of first order transitions, 
which starts at zero temperature, could terminate at some nonzero 
temperature, similar to the liquid-gas transition in water or nuclear 
matter.

The outline of the paper is as follows. In Section II we 
compute the full fermion propagator and discuss the
excitation spectrum in a superfluid. Then, the gap equations 
are derived in the mean-field approximation
for the model (\ref{L}). We solve them in Section III.
A logarithmic ultraviolet singularity arises, which upon renormalization
introduces a renormalization scale $\Lambda$. 
We study solutions of the gap equations
as a function of the coupling $g$ and the renormalization scale $\Lambda$.
We also analyze the temperature dependence of the gap energy and 
discuss the critical temperature $T_c$, at which superfluidity
vanishes. Section IV concludes with a summary of our results.
Appendix A contains a derivation of the grand partition function
and the gap equation in the mean-field
approximation, both for an antifermion-fermion condensate and
a fermion-fermion condensate. In
Appendix B we analyze the Dirac structure of the gap matrix
in detail and make contact to earlier work \cite{bailinlove}.
In Appendix C we discuss the renormalization of the gap equations.
Our units are $\hbar=k_B=c=1$. The metric tensor is $g^{\mu \nu}=(+,-,-,-)$.

\section{The Gap Equations in the Ultrarelativistic Limit} 
\label{II}

A standard way to study superfluidity
is to compute the
grand partition function ${\cal Z}(T,V,\mu)$ associated with (\ref{L}) 
in the mean-field approximation
for a fermion-fermion condensate. This is done explicitly in Appendix
\ref{A}, cf.\ eq.\ (\ref{A16}). 
One arrives at an effective action of the form (cf.\
eq.\ (\ref{action}), see also eq.\ (1.1) of \cite{bailinlove})
\begin{equation} \label{effaction}
I [\bar{\psi},\psi] = \int_x \bar{\psi}(x)
\left( i \gamma \cdot \partial + \mu\, \gamma_0 -m \right) \psi(x)
+ \frac{1}{2} \int_{x,y}  \left[ \bar{\psi}_C(x) \Delta(x,y) 
\psi(y) + {\rm h.c.} \right]\,\, ,
\end{equation}
where $\psi_C(x)$ is the charge-conjugate spinor, defined by
\begin{equation} \label{conjugatespinor}
\psi_C(x) = C \, \bar{\psi}^T(x)\,\,\, , \,\,\,\, 
\bar{\psi}_C(x) = \psi^T(x) \, C \,\,\, , \,\,\,\, 
\psi(x) = C \, \bar{\psi}_C^T(x) \,\,\, , \,\,\,\, 
\bar{\psi}(x) = \psi_C^T(x) \,C \,\, ,
\end{equation}
and $C = i \gamma^2 \gamma_0$ is the charge-conjugation matrix in Dirac 
representation, $C = - C^{-1}= - C^T = -C^\dagger$.
We also abbreviated $\int_x \equiv \int_0^{1/T} 
{\rm d} (it) \int_V {\rm d}^3 {\bf x}$.
$\Delta(x,y)$ is the gap matrix.
If the system is translationally invariant, $\Delta(x,y) \equiv
\Delta(x-y)$, the Fourier transform
\begin{equation} \label{4}
\Delta (k) = \int_x \, e^{i k \cdot x}\, \Delta(x)
\end{equation}
obeys the {\em gap equation} (cf.\ eq.\ (\ref{gapeq2}), see also
eq.\ (1.34) of \cite{bailinlove})
\begin{equation} \label{gapequation}
\Delta(k) = g^2 \frac{T}{V} \sum_q 
D(k-q,M_s)\,  G_0^{-}(q) \, \Delta(q) \, G^{+}(q) \,\, .
\end{equation}
Here, $k_0 = -i(2n+1) \pi T$,
$\sum_k \equiv \sum_n V \int {\rm d}^3 {\bf k} / (2 \pi)^3$, and
\begin{equation}
\left[G_0^{\pm}\right]^{-1} \!(k) \equiv \gamma \cdot k \pm \mu \, 
\gamma_0 -m 
\end{equation}
is the free inverse fermion propagator, while 
\begin{equation} \label{7}
G^{\pm} \equiv \left\{ \left[ G_0^{\pm} \right]^{-1} - \Sigma^{\pm} 
\right\}^{-1} 
\end{equation}
is the fermion propagator dressed by the interaction with the
fermion-fermion condensate,
$\Sigma^{\pm} \equiv \Delta^{\mp}\, G_0^{\mp}\, \Delta^{\pm}, \,
\Delta^{+} \equiv \Delta, \,\Delta^{-} \equiv \gamma_0 \Delta^\dagger
\gamma_0$.
$G_0^+$ corresponds to propagation of free particles, $G_0^-$ to
that of charge-conjugate particles. $G^+$ propagates
quasiparticles and $G^-$ charge-conjugate quasiparticles, respectively.
$D(p,M_s) = 1/(M_s^2 - p^2)$ is the propagator of the scalar boson.

In Appendix \ref{B} we show that massive fermions have eight
possible gaps in the $J=0$ channel.  In the ultrarelativistic limit,
these reduce to four.  They correspond to the pairing of
right-handed fields with positive helicity with themselves, {\it etc.}
Consequently, the gap matrix (\ref{4}) has the form
\begin{equation} \label{URlimit}
\Delta(k)  = \phi_{r +}^{+}(k) \, {\cal P}_{r +}^{+} ({\bf k})
        + \phi_{\ell -}^{+}(k) \, {\cal P}_{\ell -}^{+}({\bf k})
        + \phi_{r -}^{-}(k) \, {\cal P}_{r -}^{-}({\bf k})
        + \phi_{\ell +}^{-}(k) \, {\cal P}_{\ell +}^{-}({\bf k}) \,\, .
\end{equation}
Here,
\begin{equation}
{\cal P}_{r+}^+ ({\bf k}) \equiv {\cal P}_{r} \,
{\cal P}_+ ({\bf k}) \,\,\,\, , \,\,\,\,
{\cal P}_{\ell-}^+ ({\bf k}) \equiv {\cal P}_{\ell} \,
{\cal P}_- ({\bf k}) \,\,\,\,, \,\,\,\,
{\cal P}_{r-}^- ({\bf k}) \equiv {\cal P}_{r} \,
{\cal P}_- ({\bf k}) \,\,\,\, , \,\,\,\,
{\cal P}_{\ell+}^- ({\bf k}) \equiv {\cal P}_{\ell} \,
{\cal P}_+ ({\bf k}) \,\,, \label{9}
\end{equation}
are projectors onto states with given chirality and helicity, as denoted
by the two subscripts, where
\begin{equation} \label{10}
{\cal P}_{r,\ell} \equiv \frac{1 \pm \gamma_5}{2} \,\,\, \, , \,\,\,\,
{\cal P}_\pm ({\bf k}) \equiv \frac{1 \pm \gamma_5 \gamma_0 \bbox{\gamma}
\cdot \hat{\bf k}}{2} \,\, .
\end{equation}
The additional superscript in (\ref{9}) refers to the sign of the
energy for a non-interacting fermion field.
For massless fermions, this superscript is in principle superfluous: 
right- (left-) handed fermions with positive (negative) helicity 
{\em must\/} have positive energy, 
while those with negative (positive) helicity {\em must\/}
have negative energy.
Nevertheless, we find it physically illuminating to retain it.

The quantities $\phi_{r, \ell\, \pm}^\pm$ are the individual
gap functions; for example, $\phi_{r +}^+$ corresponds to
the condensation of two right-handed particles with positive
helicity, while $\phi_{r -}^-$ corresponds to
the condensation of two right-handed antiparticles with negative
helicity.  
This means that in the $J=0$ channel,
fermions of {\em different\/} helicity or chirality do not 
condense.  This is due to 
the projectors in eq.\ (\ref{URlimit}):
the effective action (\ref{effaction}) reduces to a sum of
four terms, one for each given chirality and helicity, see eq.\ (\ref{B34}).

The fact that only fermions with the same helicity condense in the
$J=0$ channel is physically obvious in the center-of-momentum frame 
of the pair, and holds for either massive or massless fermions.
In that frame, the requirement of equal helicity means that the 
projections of the spin along the direction of motion are
antiparallel, as in the usual non-relativistic
treatment of superconductivity.
Taking the fermions to move in the $z$ direction,
in the center-of-momentum frame
$L_z = 0$; thus if $J=0$, also $S_z =0$.
The spin wave function with $S_z =0$ can be either
the antisymmetric singlet, $S=0$, or the symmetric triplet, $S=1$.
To obtain total spin $J=0$, for the former, $S=0$ combines with
$L=0$, while for the latter, $S=1$ combines with $L=1$.
The signal for $L=1$ is the appearance of 
$\hat{\bf k}$ in eqs.\ (\ref{URlimit}), (\ref{10}).

In the following, we first discuss the structure of the full
fermion propagator (\ref{7}), because the poles of the full propagator
determine the excitation spectrum in a superfluid.
We then return to the gap equation (\ref{gapequation}).

\subsection{The full fermion propagator}

In the ultrarelativistic limit, 
the full propagator (\ref{7}) assumes the form
\begin{equation} \label{7A}
G^+(q) = \left[\gamma \cdot q + \mu \gamma_0
- \gamma_0 
\left[\Delta(q)\right]^\dagger \gamma_0 
\left(\gamma \cdot q- \mu \gamma_0\right)^{-1} \Delta(q) \right]^{-1} \,\, .
\end{equation}
From eq.\ (\ref{URlimit}) one computes
\begin{equation}
\gamma_0 \,\Delta^\dagger \, \gamma_0
= \left[\phi_{r+}^{+} \right]^\dagger {\cal P}_{\ell +}^{-} 
+ \left[\phi_{\ell -}^{+} \right]^\dagger {\cal P}_{r -}^{-} 
+ \left[\phi_{r -}^{-} \right]^\dagger {\cal P}_{\ell -}^{+} 
+ \left[\phi_{\ell +}^{-} \right]^\dagger {\cal P}_{r +}^{+} \,\,.
\end{equation}
Since the condensates $\phi$ are not matrices (i.e., unlike the QCD case,
they do not carry other internal degrees of freedom like color and flavor), 
the hermitean conjugation is replaced by simple complex conjugation. 
With the identities
\begin{equation} \label{identity}
\gamma_0 \Delta^\dagger \gamma_0 
\left(\gamma \cdot q- \mu \gamma_0\right) \Delta
= \left(\gamma \cdot q- \mu \gamma_0\right) 
\left( |\phi_{r +}^{+}|^2\, {\cal P}_{r +}^{+} 
     + |\phi_{\ell -}^{+}|^2 \, {\cal P}_{\ell -}^{+}
     + |\phi_{r -}^{-}|^2 \, {\cal P}_{r -}^{-} 
     + |\phi_{\ell +}^{-}|^2 \, {\cal P}_{\ell +}^{-}
\right)
\end{equation}
and
\begin{equation}
\left(\gamma \cdot q - \mu \gamma_0 \right)
\left(\gamma \cdot q + \mu \gamma_0 \right)
 =  \left[q_0^2 - (|{\bf q}|-\mu)^2 \right]\, \left({\cal P}_{r+}^+
+{\cal P}_{\ell -}^+ \right) 
+ \left[q_0^2 - (|{\bf q}|+\mu)^2 \right]\, \left({\cal P}_{r-}^-
+{\cal P}_{\ell +}^-\right) \,\, ,
\end{equation}
one derives
\begin{equation}
G^+(q) = \left[
\frac{{\cal P}_{r +}^{+}({\bf q})}{q_0^2 - 
\left[\epsilon^+(\phi_{r +}^{+})\right]^2}  
+ \frac{{\cal P}_{\ell -}^{+}({\bf q})}{q_0^2 - 
\left[\epsilon^+(\phi_{\ell -}^{+}) \right]^2} 
+\, \frac{ {\cal P}_{r -}^{-}({\bf q})}{q_0^2- 
\left[\epsilon^-(\phi_{r -}^{-})\right]^2} 
+ \frac{{\cal P}_{\ell +}^{-}({\bf q})}{q_0^2- 
\left[\epsilon^-(\phi_{\ell +}^{-})  \right]^2} \,  \right]\, 
 (\gamma \cdot q - \mu \gamma_0) \,\, , \label{fullprop}
\end{equation}
where 
\begin{equation}
\epsilon^\pm (\phi) \equiv \left[ (|{\bf q}| \mp \mu)^2 + 
|\phi|^2 \right]^{1/2} \,\, .
\end{equation}
The full propagator (\ref{fullprop}) has eight poles in the 
complex $q_0$ plane:
\begin{equation} \label{poles}
q_0^{(1,2)} = \pm \epsilon^+(\phi_{r+}^{+}) \,\,\,\, , \,\,\,\,
q_0^{(3,4)} = \pm \epsilon^+(\phi_{\ell -}^{+}) \,\,\,\, , \,\,\,\,
q_0^{(5,6)} = \pm \epsilon^-(\phi_{r-}^{-}) \,\,\,\, , \,\,\,\,
q_0^{(7,8)} = \pm \epsilon^-(\phi_{\ell +}^{-})\,\, .
\end{equation}
These poles deserve further discussion. 
The kinetic energy of non-interacting, massless particles 
with 3-momentum ${\bf q}$ is $|{\bf q}|$.
Let $\omega(|{\bf q}|)$ denote the energy of the quasiparticle excitations
as a function of the kinetic energy of the non-interacting particles.
In a non-interacting, ultrarelativistic Fermi system
quasiparticles (quasi-antiparticles) are the usual 
particles (antiparticles) and consequently
the excitation spectrum is
\begin{equation}
\begin{array}{lll}
\omega_0(|{\bf q}|) - \mu  & = +\, |{\bf q}| -\mu  &
\hspace*{1cm} \mbox{\rm particles} \,\, , \\
 & =  -\, |{\bf q}| +\mu&
\hspace*{1cm} \mbox{\rm particle holes} \,\, , \\
 & =  -\, |{\bf q}|-\mu  &
\hspace*{1cm} \mbox{\rm antiparticles} \,\, , \\
 & =  +\, |{\bf q}| +\mu &
\hspace*{1cm} \mbox{\rm antiparticle holes} \,\, .
\end{array}
\end{equation}
On the other hand, in a superfluid the pairing of particles and antiparticles
changes the excitation spectrum. As is well-known \cite{fetterwalecka},
the excitations are now quasiparticles (quasi-antiparticles) which
are in essence a linear superposition of particles (antiparticles)
and particle holes (antiparticle holes). This linear superposition is
the well-known Bogoliubov transformation \cite{fetterwalecka}. 
The quasiparticle excitations are sometimes called ``Bogoliubons''. 
On account of (\ref{poles}), in a superfluid 
the spectrum of right-handed excitations is
\begin{equation} \label{righthand}
\begin{array}{lll}
\omega(|{\bf q}|) - \mu  & \equiv  - \, \epsilon^+(\phi_{r+}^+) =  
- \, \sqrt{(|{\bf q}|-\mu)^2+|\phi_{r+}^+|^2} & 
\hspace*{1cm} \mbox{\rm right-handed quasiparticles} \,\, , \\
 & \equiv + \, \epsilon^+(\phi_{r+}^+) =  
+ \, \sqrt{(|{\bf q}|-\mu)^2+|\phi_{r+}^+|^2} &  
\hspace*{1cm} \mbox{\rm right-handed quasiparticle holes} \,\, , \\
 & \equiv - \, \epsilon^-(\phi_{r-}^-) = 
- \, \sqrt{(|{\bf q}|+\mu)^2+|\phi_{r-}^-|^2} & 
\hspace*{1cm} \mbox{\rm right-handed quasi-antiparticles} \,\, , \\
 & \equiv + \, \epsilon^-(\phi_{r-}^-) =  
+ \, \sqrt{(|{\bf q}|+\mu)^2+|\phi_{r-}^-|^2}  & 
\hspace*{1cm} \mbox{\rm right-handed quasi-antiparticle holes} \,\, ,
\end{array}
\end{equation}
while for left-handed excitations one has
\begin{equation} \label{lefthand}
\begin{array}{lll}
\omega(|{\bf q}|) - \mu  & \equiv - \, \epsilon^+(\phi_{\ell -}^+) =  
- \, \sqrt{(|{\bf q}|-\mu)^2+|\phi_{\ell-}^+|^2} & 
\hspace*{1cm} \mbox{\rm left-handed quasiparticles} \,\, , \\
 & \equiv + \, \epsilon^+(\phi_{\ell-}^+) =  
+ \, \sqrt{(|{\bf q}|-\mu)^2+|\phi_{\ell -}^+|^2} &  
\hspace*{1cm} \mbox{\rm left-handed quasiparticle holes} \,\, , \\
 & \equiv - \, \epsilon^-(\phi_{\ell+}^-) = 
- \, \sqrt{(|{\bf q}|+\mu)^2+|\phi_{\ell+}^-|^2} & 
\hspace*{1cm} \mbox{\rm left-handed quasi-antiparticles} \,\, , \\
 & \equiv + \, \epsilon^-(\phi_{\ell+}^-) =  
+ \, \sqrt{(|{\bf q}|+\mu)^2+|\phi_{\ell +}^-|^2}  & 
\hspace*{1cm} \mbox{\rm left-handed quasi-antiparticle holes} \,\, .
\end{array}
\end{equation}

\begin{figure} 
\hspace*{2.5cm} 
\psfig{figure=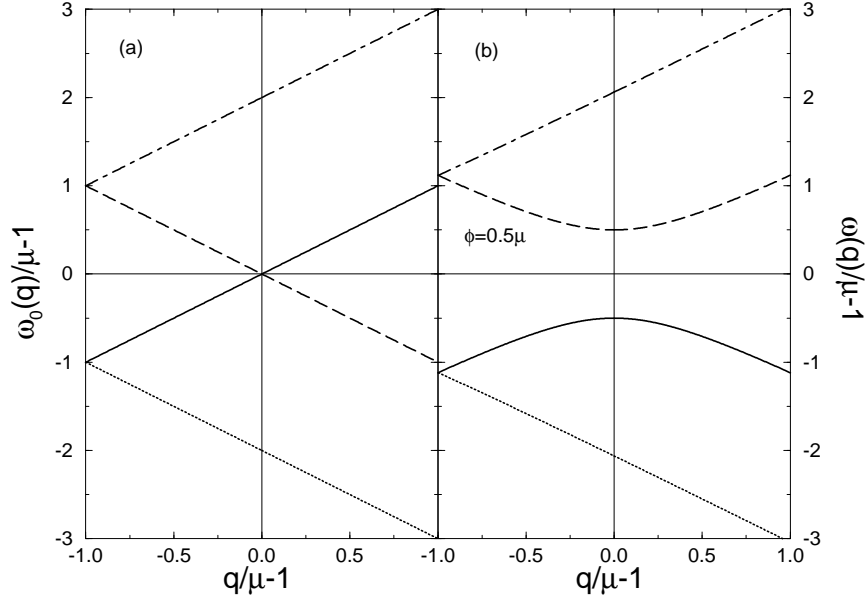,width=2.7in,height=3.5in,angle=-90}
\vspace*{-1cm}
\caption{The dispersion relation for
(quasi)particle (solid), (quasi-)antiparticle (dotted), 
(quasi)particle-hole (dashed),
and (quasi-)antiparticle-hole states (dot-dashed). (a) Non-interacting,
massless fermions, (b) superfluid, massless fermions.}
\label{dispersion}
\end{figure}

These branches are displayed in Fig.\ \ref{dispersion}, for simplicity
with a common gap $\phi=0.5\, \mu$.
For free fermions, the particle and hole branches cross at the
Fermi surface. The generation of a gap produces a phenomenon analogous to
level repulsion in quantum mechanics, as the quasiparticle and 
quasiparticle-hole branches become disconnected. 
Relative to the non-interacting case, the meaning of quasiparticle
and quasiparticle-hole branch interchanges above the Fermi surface.

The main difference from the non-relativistic case is the appearance
of the (quasi-)antiparticle branch and the (quasi-)antiparticle-hole branch. 
This means that condensation is not only restricted to particles 
close to the Fermi surface,
but that also antiparticles in the Dirac sea condense.
This can be confirmed by considering the fermion-fermion scattering
amplitude: a BCS-type singularity occurs not only for particle-particle
scattering, but also for antiparticle-antiparticle scattering, giving
rise to the antiparticle condensates $\phi^-_{r-,\ell+}$.
Notice, however,
that it is always easier to excite a quasiparticle than a quasi-antiparticle.
It costs only an energy $2\, \phi$ to excite a quasiparticle at the Fermi
surface, but $2 \sqrt{\mu^2 + \phi^2}$ to excite a
quasi-antiparticle with zero momentum.
\newpage
\vspace*{0.5cm}

One can also compute the occupation numbers for quasiparticles and
quasi-antiparticles. To this end, it is advantageous to
Fourier-transform the full propagator to the so-called
``mixed'' (or ``Saclay'') representation
\begin{equation} \label{saclayrep}
G^+(\tau,{\bf q}) = T \sum_n e^{i \omega_n \tau} \, G^+(\omega_n,{\bf q})
\,\, ,
\end{equation}
where $i\omega_n = i (2 n+1) \pi T \equiv - q_0$.
For instance, the free propagator for massless fermions reads
in this representation \cite{LeBellac}
\begin{eqnarray}
G_0^+(\tau,{\bf q}) & = & - \left[ \theta(\tau) - N_F\left( 
\frac{|{\bf q}|- \mu}{T} \right) \right]  \exp[-(|{\bf q}|- \mu)\tau]
\, \Lambda^+({\bf q}) \gamma_0 \nonumber \\
&   & + \left[ \theta(-\tau) - N_F\left( 
\frac{|{\bf q}|+ \mu}{T} \right) \right]  \exp[(|{\bf q}|+ \mu)\tau] 
\, \Lambda^-({\bf q}) \gamma_0\,\, , \label{free}
\end{eqnarray}
where $N_F(x) \equiv (e^x + 1)^{-1}$ is the usual Fermi--Dirac
distribution and 
$\Lambda^\pm({\bf q})$ is the projector onto positive (negative)
energy states (eq.\ (\ref{energyproj}) in the massless limit, i.e.,
for $\beta_{\bf q}=1$ and $\alpha_{\bf q}=0$).
The occupation numbers are now directly exhibited as the prefactors of the 
exponential (imaginary) time evolution factors. For $\tau \geq 0$, the
first term corresponds to the propagation of particle holes, with
the thermal occupation number $1-N_F[(|{\bf q}|- \mu)/T]$, while
the second term corresponds to the propagation of antiparticles, with
the thermal occupation number $N_F[(|{\bf q}|+ \mu)/T]$.
For $\tau <0$, the first term corresponds to the propagation of
particles, with the thermal occupation number $N_F[(|{\bf q}|- \mu)/T]$.
The second corresponds to the propagation of antiparticle holes,
with occupation number $1-N_F[(|{\bf q}|+ \mu)/T]$.

The full fermion propagator is computed from eq.\ (\ref{saclayrep}) with
(\ref{fullprop}). The result is
\begin{eqnarray}
G^+(\tau,{\bf q}) & = & - \sum_{hs = r+, \ell -} \left( 
\left\{ \theta(\tau) - 
N_F\left[\frac{\epsilon^+(\phi_{hs}^+)}{T}\right] \right\}
\left[ 1 - n_{\bf q}^+(\phi_{hs}^+) \right]
\exp\left[-\epsilon^+(\phi_{hs}^+) \tau \right]
\right. \nonumber \\
&   & \hspace*{1.4cm} - \left. \left\{ \theta(-\tau) - 
N_F\left[\frac{\epsilon^+(\phi_{hs}^+)}{T}\right] \right\}
\, n_{\bf q}^+(\phi_{hs}^+) \, \exp\left[\epsilon^+(\phi_{hs}^+) \tau \right] 
\right) {\cal P}^+_{hs}({\bf q}) \gamma_0 \nonumber \\
&  &  - \sum_{hs = r-, \ell +} \left(
\left\{ \theta(\tau) - 
N_F\left[\frac{\epsilon^-(\phi_{hs}^-)}{T}\right] \right\}
\left[ 1 - n_{\bf q}^-(\phi_{hs}^-) \right]
\exp\left[-\epsilon^-(\phi_{hs}^-) \tau \right]
\right. \nonumber \\
&   & \hspace*{1.4cm} - \left. \left\{ \theta(-\tau) - 
N_F\left[\frac{\epsilon^-(\phi_{hs}^-)}{T}\right] \right\}
\, n_{\bf q}^-(\phi_{hs}^-) \, \exp\left[\epsilon^-(\phi_{hs}^-) \tau \right] 
\right) {\cal P}^-_{hs}({\bf q}) \gamma_0 \,\, , \label{saclay}
\end{eqnarray}
where 
\begin{equation} \label{occno}
n_{\bf q}^\pm(\phi) \equiv \frac{\epsilon^\pm(\phi) \mp (|{\bf q}|\mp \mu)}{
2 \, \epsilon^\pm(\phi)} \,\, .
\end{equation}
In deriving eq.\ (\ref{saclay}), in an intermediate step one
encounters terms of the form
$$
\frac{(\epsilon^\pm - \mu)\, \gamma_0 - \bbox{\gamma} \cdot {\bf q}}{2\,
\epsilon^\pm} \hspace*{1cm} {\rm and} \hspace*{1cm}
\frac{-(\epsilon^\pm + \mu)\, \gamma_0 - \bbox{\gamma} \cdot {\bf q}}{2\,
\epsilon^\pm}\,\, .
$$
By adding and subtracting a term $|{\bf q}|\, \gamma_0$ in the
numerator, one can transform them into
$$
\gamma_0 \, \left(1-n_{\bf q}^\pm\right) \mp \Lambda^\mp({\bf q}) \,
\gamma_0 \, \frac{|{\bf q}|}{\epsilon^\pm} \hspace*{1cm} {\rm and} 
\hspace*{1cm}
- \gamma_0 \, n_{\bf q}^\pm \mp \Lambda^\mp({\bf q}) \,
\gamma_0 \, \frac{|{\bf q}|}{\epsilon^\pm} \,\, .
$$
Eq.\ (\ref{saclay}) then follows by making use of eq.\ (\ref{B31})
and the orthogonality of the energy projectors, 
$\Lambda^+ \Lambda^- \equiv 0$.

There is a one-to-one correspondence between the eight poles of the full
propagator (\ref{poles}) and the eight terms on the right-hand side
of eq.\ (\ref{saclay}). The first line in eq.\ (\ref{saclay}) arises from
the pole for right(left)-handed quasiparticle holes, the second line from
that for right(left)-handed quasiparticles, while the third line
originates from the pole for right(left)-handed quasi-antiparticle holes
and the fourth line from that for right(left)-handed quasi-antiparticles.
As in the non-interacting case (\ref{free}), the occupation numbers can 
now be directly read off as the prefactors of the exponential 
time evolution factors. 

\begin{figure} 
\hspace*{2.5cm} 
\psfig{figure=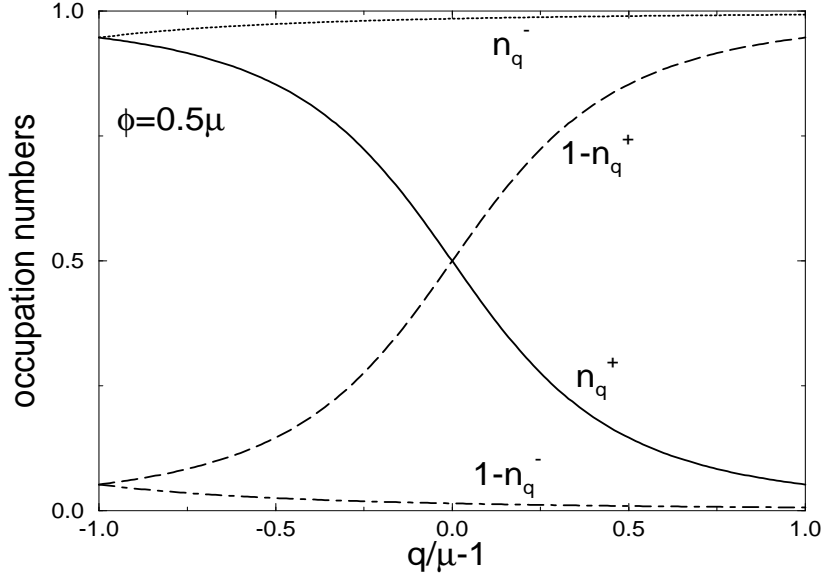,width=2.7in,height=3.3in,angle=-90}
\vspace*{-1cm}
\caption{The occupation number for
quasiparticles $n_{\bf q}^+$ (solid), quasi-antiparticles 
$n_{\bf q}^-$ (dotted), quasiparticle holes $1-n_{\bf q}^+$ (dashed),
and quasi-antiparticle holes $1-n_{\bf q}^-$ (dot-dashed) 
in a superfluid at $T=0$.}
\label{figoccno}
\end{figure}

It is interesting to study two limiting cases of (\ref{saclay}).
The first is the zero-temperature limit, where
the Fermi--Dirac distributions vanish on account of the fact
that their argument is always positive.
Then, the functions $n_{\bf q}^\pm$ defined in eq.\ (\ref{occno}) are
identified with the {\em occupation numbers\/} for quasiparticles and
quasi-antiparticles in a superfluid at $T=0$. Correspondingly,
$1-n_{\bf q}^\pm$ are the occupation numbers of quasiparticle holes
and quasi-antiparticle holes, respectively. These are shown in
Fig.\ \ref{figoccno}. The quasiparticle and quasiparticle-hole 
occupation numbers exhibit the smearing around the Fermi surface
characteristic for a superfluid or a superconductor
\cite{fetterwalecka}. The smearing is a consequence of the fact
that the ``Bogoliubons'' are superpositions of particle and hole
states. Note also that there are always some quasi-antiparticle-hole
excitations present.

\vspace*{1cm}

The other interesting limit is the non-interacting case where one
is supposed to recover eq.\ (\ref{free}).
Taking all gaps to vanish, $\phi_{hs}^\pm \rightarrow 0$, one has
$\epsilon^\pm \rightarrow | |{\bf q}| \mp \mu|$, such that
\begin{equation}
n_{\bf q}^+ \rightarrow \theta(\mu-|{\bf q}|) \,\,\,\, , \,\,\,\,
n_{\bf q}^- \rightarrow 1 \,\,\,\, (\phi \rightarrow 0)\,\, .
\end{equation}
The contribution from quasi-antiparticle holes vanishes completely, and
the quasi-antiparticle term becomes the antiparticle term,
the second line of eq.\ (\ref{free}).
However, in order to obtain the particle-hole contribution (the first line
of eq.\ (\ref{free})), one has to {\em combine\/} the
contribution from quasiparticles and quasiparticle holes.
This can be understood by inverting the Bogoliubov transformation which
combines particles and particle holes to
quasiparticles (and quasiparticle holes).

\subsection{The gap equations}

After discussing the full fermion propagator and the
excitation spectrum in a superfluid, we now return to
the discussion of the gap equation (\ref{gapequation}) in the
ultrarelativistic limit. With the expression (\ref{fullprop})
for the full propagator, it assumes the form
\begin{eqnarray} 
\Delta(k) & = & g^2 \frac{T}{V} \sum_q D(k-q,M_s) \, \left[
\frac{\phi_{r +}^{+}(q)}{q_0^2 - \left[\epsilon^+(\phi_{r +}^{+})\right]^2}  
    \, {\cal P}_{\ell +}^{-}({\bf q}) 
+ \frac{\phi_{\ell -}^{+}(q)}{q_0^2 - \left[\epsilon^+(\phi_{\ell -}^{+})
 \right]^2} \, {\cal P}_{r -}^{-}({\bf q})  \right.  \nonumber \\
&   & \left. \hspace*{3.3cm}
+\, \frac{\phi_{r -}^{-}(q)}{q_0^2- \left[\epsilon^-(\phi_{r -}^{-})\right]^2} 
   \, {\cal P}_{\ell -}^{+}({\bf q}) 
+ \frac{\phi_{\ell +}^{-}(q)}{q_0^2- \left[\epsilon^-(\phi_{\ell +}^{-})
  \right]^2} \, {\cal P}_{r +}^{+}({\bf q}) \right] \,\, .\label{41}
\end{eqnarray}
Forming suitable projections of eq.\ (\ref{URlimit}), one derives 
gap equations for the individual gap functions:
\begin{mathletters}
\begin{eqnarray}
\phi_{r+}^{+} (k) & = & \frac{g^2}{2} \frac{T}{V} \sum_q \,D (k-q,M_s) \left[
\frac{1 - \hat{\bf k} \cdot \hat{\bf q}}{q_0^2-
 \left[\epsilon^+(\phi_{\ell -}^{+})\right]^2}  \, \phi_{\ell -}^{+}(q)
+\frac{1 + \hat{\bf k} \cdot \hat{\bf q}}{q_0^2-
 \left[\epsilon^-(\phi_{\ell +}^{-})\right]^2} \, \phi_{\ell +}^{-}(q) 
\right] \,\, , \\
\phi_{\ell-}^{+}(k) & = & \frac{g^2}{2}\frac{T}{V} \sum_q \,D (k-q,M_s)\left[
\frac{1 - \hat{\bf k} \cdot \hat{\bf q}}{q_0^2-
 \left[\epsilon^+(\phi_{r+}^{+})\right]^2}  \, \phi_{r+}^{+}(q)
+\frac{1 + \hat{\bf k} \cdot \hat{\bf q}}{q_0^2-
 \left[\epsilon^-(\phi_{r-}^{-})\right]^2} \,  \phi_{r-}^{-}(q) 
\right] \,\, , \\
\phi_{r-}^{-} (k) & = & \frac{g^2}{2} \frac{T}{V} \sum_q \,D (k-q,M_s) \left[
\frac{1 + \hat{\bf k} \cdot \hat{\bf q}}{q_0^2 -
 \left[\epsilon^+(\phi_{\ell -}^{+})\right]^2} \,  \phi_{\ell -}^{+}(q)
+ \frac{1 - \hat{\bf k} \cdot \hat{\bf q}}{q_0^2-
 \left[\epsilon^-(\phi_{\ell +}^{-})\right]^2} \,  \phi_{\ell +}^{-}(q) 
\right] \,\, , \\
\phi_{\ell +}^{-}(k) & = & \frac{g^2}{2}\frac{T}{V} \sum_q \,D (k-q,M_s)\left[
\frac{1 + \hat{\bf k} \cdot \hat{\bf q}}{q_0^2 -
 \left[\epsilon^+(\phi_{r+}^{+})\right]^2} \,    \phi_{r+}^{+}(q)
+ \frac{1 - \hat{\bf k} \cdot \hat{\bf q}}{q_0^2 -
 \left[\epsilon^-(\phi_{r-}^{-})\right]^2} \,    \phi_{r-}^{-}(q) 
\right] \,\, .
\end{eqnarray} \label{fullgapeqs}
\end{mathletters}

These gap equations (\ref{fullgapeqs}) 
do {\it not\/} agree with those derived
by Bailin and Love \cite{bailinlove}.
These authors include only half of the fermion quasiparticle
modes: they take
the quasiparticle and quasiparticle-hole branches, with gaps 
$\phi_{r+}^+$ and $\phi_{\ell-}^+$, but 
neglect the quasi-antiparticle and quasi-antiparticle-hole branches, with
gaps $\phi_{r-}^-$ and $\phi_{\ell +}^-$.   
They also restrict themselves to
condensation in the $0^+$ channel, which forces $\phi_{r+}^+
= - \phi_{\ell -}^+$ (see eq.\ (\ref{0+})). 
Doing so, they find a single equation for a linear combination of
the gaps.  We do not find 
a single equation.  Instead, due to the form of the
full fermion propagator (\ref{fullprop}), each of the four
gap equations (\ref{fullgapeqs})
contains a sum of two terms: each gap on the right-hand side,  
such as $\phi_{r+}^+$, is always multiplied
by an energy denominator with {\em only\/} that gap, 
$1/\left\{q_0^2 - \left[\epsilon^+(\phi_{r+}^{+})\right]^2\right\}$.

In the case of a point-like four-fermion interaction
as in NJL-type models, $D(k-q,M_s) \rightarrow 1/M_s^2$. Then, the
terms proportional to $\hat{\bf k} \cdot \hat{\bf q}$
vanish by symmetry, and the right-hand sides of (\ref{fullgapeqs})
no longer depend on either $k_0$ or ${\bf k}$:
the gaps are simply constants. 
Defining
\begin{equation}
F(\phi) \equiv \frac{g^2}{2M_s^2}\frac{T}{V} \sum_q \left[
  \frac{1}{q_0^2 - \left[\epsilon^+(\phi)\right]^2} \, 
+ \frac{1}{q_0^2 - \left[\epsilon^-(\phi)\right]^2} \, 
\right] \,\,,
\end{equation}
the gap equations reduce to
\begin{equation} \label{fullgapeqsMsinfty2}
\phi_{r+}^+ \equiv \phi_{r-}^-
= \phi_{\ell -}^+ \, F(\phi_{\ell -}^+) \,\,\,\, ,\,\,\,\,
\phi_{\ell -}^+ \equiv \phi_{\ell +}^- = \phi_{r+}^+ \, F(\phi_{r+}^+) \,\, .
\end{equation}
The gaps for right- and left-handed quasi-antiparticles equal those
for the corresponding quasiparticles. This reduces the number of
independent gaps to two.

The solutions of eqs.\ (\ref{fullgapeqsMsinfty2}) satisfy
$1 = F(\phi) F(\phi F(\phi))$, where $\phi$ is either 
$\phi_{r+}^+$ or $\phi_{\ell -}^+$.
One possible solution, $\phi_0$, satisfies $1 \equiv |F(\phi_0)|$;
as $F(\phi)$ is single-valued as a function of 
$\phi$ for $\phi \geq 0$, and $F(\phi) = F(-\phi)$, 
the only possible solutions are $\pm \phi_0$, therefore
$\phi_{r+}^+ = \pm \phi_{\ell -}^+$.
Since $F<0$, the solution of eqs.\ (\ref{fullgapeqsMsinfty2}) obeys
$\phi_{\ell-}^+ = - \phi_{r+}^+$. From eq.\
(\ref{0+}) we then conclude that condensation occurs
only in the $0^+$ channel.

In conclusion, for scalar NJL-type interactions there is only one
independent gap function. Physically, this happens because
for a point-like interaction there are no states
with nonzero angular momentum, and so terms in the gap
proportional to $\hat{\bf k}$ ---
which signal $L=1$ --- must vanish. 
In the next section we show that scalar boson exchange over a finite range,
$M_s < \infty$, lifts this degeneracy and produces two independent
gap functions.

\section{Solving the Gap Equations}

The set of equations (\ref{fullgapeqs})
determines the gap functions for massless
fermions in the mean-field approximation.
In principle, the gaps are functions of the 4-momentum $k^\mu$, so that
eqs.\ (\ref{fullgapeqs}) are actually {\em integral\/} equations.
In the following, we assume that 
the momentum dependence of the gap functions is negligible. 
We comment on the validity of this approximation below.  
Remember that in NJL-type models the
gap functions are automatically independent of $k^\mu$.

Under the assumption that the gap functions are simply constants,
one ends up with the following set of four {\em algebraic\/} equations:
\begin{mathletters}
\begin{eqnarray}
\phi_{r+}^{+} & = & \phi_{\ell -}^{+} \, \left[ 
        {\cal F}_0^+ (\phi_{\ell -}^{+}) -
        {\cal F}_1^+ (\phi_{\ell -}^{+})  \right]
                +   \phi_{\ell +}^{-} \, \left[ 
        {\cal F}_0^- (\phi_{\ell +}^{-}) + 
        {\cal F}_1^- (\phi_{\ell +}^{-}) \right] \,\, , \\
\phi_{\ell -}^{+} & = & \phi_{r+}^{+} \, \left[ 
        {\cal F}_0^+ (\phi_{r+}^{+}) -
        {\cal F}_1^+ (\phi_{r+}^{+})  \right]
              + \phi_{r-}^{-} \, \left[ 
        {\cal F}_0^- (\phi_{r-}^{-}) + 
        {\cal F}_1^- (\phi_{r-}^{-})  \right] \,\, , \\
\phi_{r-}^{-} & = & \phi_{\ell -}^{+} \, \left[ 
        {\cal F}_0^+ (\phi_{\ell -}^{+}) + 
        {\cal F}_1^+ (\phi_{\ell -}^{+}) \right]
              + \phi_{\ell +}^{-} \, \left[ 
        {\cal F}_0^- (\phi_{\ell +}^{-}) - 
        {\cal F}_1^- (\phi_{\ell +}^{-}) \right] \,\, , \\
\phi_{\ell +}^{-} & = & \phi_{r+}^{+} \, \left[ 
        {\cal F}_0^+ (\phi_{r+}^{+})  + 
        {\cal F}_1^+ (\phi_{r+}^{+}) \right]
              + \phi_{r-}^{-} \, \left[ 
        {\cal F}_0^- (\phi_{r-}^{-}) - 
        {\cal F}_1^- (\phi_{r-}^{-}) \right] \,\, , 
\end{eqnarray} \label{algebraicgapeqs}
\end{mathletters}
where
\begin{equation} \label{EF}
{\cal F}_n^{\pm}(\phi) \equiv \frac{g^2}{2} \, \frac{T}{V} \sum_q
D(k-q,M_s) \, \frac{(\hat{\bf k} \cdot \hat{\bf q})^n}{q_0^2 - 
\left[\epsilon^\pm(\phi)\right]^2} 
\,\, .
\end{equation}
Note that ${\cal F}_1^\pm \equiv 0$ for NJL-type models.

The functions ${\cal F}_{0,1}^\pm$ are further evaluated 
replacing the Matsubara sum over $q_0 \equiv - i (2n+1) \pi T$ by a contour
integral and applying the residue theorem. Poles in the complex
$q_0$ plane arise from the fermion as well as the boson propagator.
Keeping only the former, we obtain
\begin{mathletters} \label{F01pm}
\begin{eqnarray} \label{F0pm}
{\cal F}_0^{\pm} & \equiv & - \frac{g^2}{32\pi^2 k} \int_0^\infty
{\rm d}q \,  \ln \left[ \frac{M_s^2 + (k +q)^2}{M_s^2 + (k -q)^2}
\right] \,
\frac{q }{\epsilon^\pm} \, \tanh \, 
\left[ \frac{\epsilon^\pm}{2T} \right] \,\, , \\
{\cal F}_1^{\pm}  & \equiv & - \frac{g^2}{32\pi^2 k} \int_0^\infty
{\rm d}q \, \left\{ \frac{M_s^2 + k^2 + q^2}{2 q k} \,
\ln \left[ \frac{M_s^2 + (k +q)^2}{M_s^2 + (k -q)^2}\right] -2 \right\}\,
\frac{q}{\epsilon^\pm} \, \tanh \, 
\left[ \frac{\epsilon^\pm}{2T} \right] \,\, ,
\end{eqnarray}
\end{mathletters}
where we have performed the angular integration and denoted 
$k \equiv |{\bf k}|,\, q \equiv |{\bf q}|$.
Since ${\cal F}_{0,1}^\pm$ are real, the gap functions $\phi$ 
can be chosen to be real.

To obtain this result we assume that the exchanged boson has
zero energy, $q_0 = k_0$. This approximation can be justified
as follows. The dominant term in 
${\cal F}_{0,1}^+$ comes from particles close to the
Fermi surface, $q \simeq \mu$. Assuming that $|\phi| \ll \mu$, this produces a
logarithmic dependence on $|\phi|$:
\begin{equation} \label{log}
\int_0^{\Lambda_{\rm UV}} \frac{{\rm d} q}{\epsilon^+} = 
\int_0^{\Lambda_{\rm UV}}
\frac{{\rm d}q}{\sqrt{(q-\mu)^2 + |\phi|^2}} \sim 
\ln \frac{\Lambda_{\rm UV}}{|\phi|} \,\, .
\end{equation}
(We discuss the ultraviolet cut-off $\Lambda_{\rm UV} \gg \mu$ below.)
The assumption that the gap $|\phi|$ is small relative
to the Fermi energy is justified in weak coupling, $g \ll 1$. 
To obtain the logarithm (\ref{log}), the fermions have to stay close to
the Fermi surface. This can only be achieved if the exchanged boson has
zero energy, $q_0 = k_0$. 

Poles of the boson
propagator, $q_0 = k_0 \pm \sqrt{M_s^2 + ({\bf k} - {\bf q})^2}$,
which we neglected in deriving eqs.\ (\ref{F01pm}), represent
processes in which the fermions in the gap equation are far from
the Fermi surface; this produces terms of order 1, but not
of order $\ln (\Lambda_{\rm UV}/|\phi|)$. (Further, keeping only zero-energy
bosons also eliminates a possible dependence of ${\cal F}_{0,1}^\pm$
on $k_0$.)

Analogous to eq.\ (\ref{log}), the functions ${\cal F}_{0,1}^-$ behave as
\begin{equation}
\int_0^{\Lambda_{\rm UV}} \frac{{\rm d} q}{\epsilon^-} = 
\int_0^{\Lambda_{\rm UV}}
\frac{{\rm d}q}{\sqrt{(q+\mu)^2 + |\phi|^2}} \sim 
\ln \frac{\Lambda_{\rm UV}}{\mu} \,\,
\end{equation}
in the limit $\Lambda_{\rm UV} \gg \mu  \gg |\phi|$.
Therefore, ${\cal F}_{0,1}^-$ do not include terms 
$\sim \ln (\Lambda_{\rm UV}/|\phi|)$ and, in weak coupling, can be neglected
relative to ${\cal F}_{0,1}^+$. Physically, this is because
$\epsilon^-$ represents the excitation spectrum of quasi-antiparticles
which are always far from the Fermi surface, cf.\ Fig.\ \ref{dispersion}.

While our approximations are controlled only in weak coupling,
we nevertheless find it illustrative to consider the qualitative
nature of the solutions in strong coupling.
For $g \simeq 1$, the gap can be of order $\mu$. 
Similarly, the functions  ${\cal F}_{0,1}^-$ are of comparable
magnitude to ${\cal F}_{0,1}^+$. We therefore retain them
in the following analysis.

It is surprising that an ultraviolet cut-off $\Lambda_{\rm UV}$ appears in
the gap equations. There is no ultraviolet divergence for
the gap equations of ordinary superconductors \cite{fetterwalecka}, as
a cut-off is provided by the Debye frequency, $\omega_D \ll \mu$, so that
the integrals receive contributions only from a
narrow interval around the Fermi surface, $\mu - \omega_D \leq q
\leq \mu + \omega_D$. On the other hand, in $He$--3 the cut-off is provided
by the chemical potential, $\mu$.

The appearance of $\Lambda_{\rm UV}$ is an artefact of
our approximation of taking a constant gap. This approximation
is manifestly inconsistent when $k \gg \mu$.
The gap functions fall off at large $k$, removing the 
apparent ultraviolet divergence. Consequently, the true gap function is
proportional to $\mu$, not $\Lambda_{\rm UV}$. That there is no
ultraviolet divergence in the true gap function can also be seen
by considering the particle-particle and antiparticle-antiparticle scattering 
amplitudes. The box diagrams whose infrared singularities generate
a non-zero gap \cite{fetterwalecka} are manifestly ultraviolet finite,
even in the relativistic regime. In order to simplify the solution
of the gap equations, however, we take the gap functions to be constant
and thus ignore these complications. 

With this approximation, 
the integral ${\cal F}_0^\pm$ is logarithmically divergent in
the ultraviolet, while ${\cal F}_1^\pm$ is finite. Instead of using a
fixed ultraviolet cut-off $\Lambda_{\rm UV}$, 
in Appendix \ref{C} we show how ${\cal F}_0^\pm$ can be rendered finite
by renormalization. The result is
\begin{equation}
{\cal F}_0^{\pm} = - \frac{g^2}{32\pi^2 k} \int_0^\infty
{\rm d}q \,\left\{ \frac{q}{\epsilon^\pm} \, 
\ln \left[ \frac{M_s^2 + (k +q)^2}{M_s^2 + (k -q)^2} \right] \, \tanh \, 
\left[ \frac{\epsilon^\pm}{2T} \right] - \ln 
\frac{(k +q)^2}{(k -q)^2} \right\} + \frac{g^2}{16\pi^2 }\ln 
\frac{k^2}{\Lambda^2 e^2}  \,\, ,
\end{equation}
where $\Lambda$ is a renormalization scale. 
One can readily convince oneself that for $\Lambda \gg k$, 
the renormalization scale can be identified with the
ultraviolet cut-off $\Lambda_{\rm UV}$ introduced above.
In light of this, we take $\Lambda \gg \mu$.

In the following, we turn to the solution of the gap equations, first
at $T=0$ and then at nonzero $T$. It is instructive to start
with the weak-coupling limit, $g \ll 1$.

\subsection{Weak-coupling limit} \label{wc}

The gap energy $\phi$ is expected to be exponentially small in 
weak coupling, $\phi \sim \mu \exp(-c/g^2)$ \cite{fetterwalecka}. 
The integrands of the functions ${\cal F}_{0,1}^+$ are
strongly peaked around $q = \mu$, since then $\epsilon^+ = |\phi| \ll \mu $.  
Consequently, the main contribution to the integrals
${\cal F}_{0,1}^+$ comes from a (small) region $\mu-\delta \leq q \leq 
\mu + \delta$, where $\delta = a \mu$, with some constant $a$
which we do not determine.
The functions ${\cal F}_{0,1}^-$ are relatively suppressed by a 
factor $\epsilon^+/\epsilon^- \simeq |\phi|/2\mu \ll 1$. 
For $\delta \gg |\phi|$:
\begin{mathletters} \label{weakcoup}
\begin{eqnarray}
{\cal F}_0^+ & \simeq & - \frac{g^2}{16\pi^2}\, \frac{\mu}{k} \,
           \ln \left[ \frac{M_s^2 + (k+ \mu)^2}{M_s^2 +(k-\mu)^2}\right] \, 
              \ln \frac{2\delta}{|\phi|}
 \,\, , \\
{\cal F}_1^+ & \simeq & - \frac{g^2}{16\pi^2}\, \frac{\mu}{k} 
   \left\{     \frac{M_s^2 + k^2 + \mu^2}{2 \mu k} \, 
     \ln\left[ \frac{M_s^2 + (k+ \mu)^2}{M_s^2 +(k-\mu)^2}\right] \,  - 2
   \right\} \, \ln \frac{2\delta}{|\phi|} \,\, , \\
{\cal F}_0^- & \simeq & 0\,\,\,\,  ,  \,\,\,\, {\cal F}_1^- \simeq 0 \,\,.
\end{eqnarray}
\end{mathletters}
Renormalization corrections are unimportant in this limit, since 
they only change $\delta$.

The gap equations (\ref{algebraicgapeqs}) were derived under
the approximation that the gaps are constants independent of $k$.
From eq.\ (\ref{weakcoup}), however, we see that 
the functions ${\cal F}_{0,1}^+$ do depend strongly on $k$, and
peak around $k=\mu$. 
Thus, as was seen previously in eq.\ (\ref{log}),
in weak coupling pairing is dominated by
fermions close to the Fermi surface. For $g \ll 1$ 
it is therefore consistent to neglect
the momentum dependence of both ${\cal F}_{0,1}^+$ and the gap 
functions and consider the above
expressions (\ref{weakcoup}) at $k= \mu$.

To one-loop order, a scalar 
mass is generated by its coupling to a fermion loop,
\begin{equation}
M_s^2 = \frac{g^2}{2}  \left(\frac{\mu^2}{\pi^2} +T^2 \right) \,\, .
\end{equation}
If we had included scalar quartic interactions in the Lagrangian
(\ref{L}), they would contribute to $M_s^2$ a term $\sim \lambda T^2$,
where $\lambda$ is the quartic coupling. Thus, at zero temperature there
is no change, while at nonzero temperature there is an innocuous
increase of $M_s^2$. As will be disussed below, this tends to
decrease $T_c$.

In weak coupling, $M_s \sim g \mu \ll \mu$,
and we can further approximate
\begin{mathletters} \label{weakcoup2}
\begin{eqnarray}
{\cal F}_0^+ & \simeq & - \frac{g^2}{16\pi^2}\, \,
              \ln\frac{4 \mu^2}{M_s^2} \, 
              \ln \frac{2\delta}{|\phi|}
 \,\, , \\
{\cal F}_1^+ & \simeq & - \frac{g^2}{16\pi^2}\, 
   \left[  \ln \frac{4 \mu^2}{M_s^2} \,  - 2
   \right] \, \ln \frac{2\delta}{|\phi|} \,\, .
\end{eqnarray}
\end{mathletters}
We can now solve the gap equations (\ref{algebraicgapeqs}).
Using (\ref{weakcoup2}) we obtain:
\begin{mathletters} \label{weakcoupgapeqs}
\begin{eqnarray}
\phi_{r+}^+ & = & - \phi_{\ell -}^+ \, \frac{g^2}{8 \pi^2} \, \ln 
                    \frac{2\delta}{|\phi_{\ell -}^+|} \,\, , \\
\phi_{\ell -}^+ & = & - \phi_{r+}^+ \, \frac{g^2}{8 \pi^2} \, \ln 
                    \frac{2\delta}{|\phi_{r+}^+|} \,\, , \\
\phi_{r-}^- & = & - \phi_{\ell -}^+ \, \frac{g^2}{8 \pi^2} \, 
                    \left[ \ln \frac{4\mu^2}{M_s^2} -1 \right]\,
                    \ln \frac{2\delta}{|\phi_{\ell -}^+|} \,\, , \\
\phi_{\ell +}^- & = & - \phi_{r+}^+ \, \frac{g^2}{8 \pi^2} \,
                    \left[ \ln \frac{4\mu^2}{M_s^2} -1 \right] \,
                  \ln \frac{2\delta}{|\phi_{r+}^+|} \,\, .
\end{eqnarray}
\end{mathletters}
By the same arguments used at the end of section \ref{II} in the
case of a point-like four-fermion interaction, the first two equations
yield $\phi_{r+}^+ \equiv - \phi_{\ell -}^+$, where $\phi_{r+}^+$
is a solution of $1 = g^2/(8\pi^2)\, \ln (2 \delta/|\phi_{r+}^+|)$.
The last two equations indicate that the gap functions for the
quasi-antiparticles are larger than those for the quasiparticles by a 
factor $\ln[4\mu^2/M_s^2]-1$. In conclusion, the solution of the
gap equations to leading order in weak coupling reads:
\begin{mathletters} \label{weakcoupresults}
\begin{eqnarray}
\phi_{r+}^+ & = & -  \phi_{\ell -}^+ = 
2\, \delta \, \exp \left[ - \frac{8 \pi^2}{g^2} \right] \sim
 \mu \, \exp \left[ - \frac{8 \pi^2}{g^2} \right] \,\,, \\
\phi_{r-}^- & = & - \phi_{\ell +}^- = 
\left[ \ln \frac{4 \mu^2}{M_s^2} - 1 \right]\, \phi_{r+}^+ \,\,.
\end{eqnarray}
\end{mathletters}
By eq.\ (\ref{0+}), this confirms the result already obtained in the limit
$M_s \rightarrow \infty$ that $J=0$ pairing of fermions interacting
via scalar boson exchange only occurs in the $0^+$ channel.
Note, however, that in contrast to the $M_s \rightarrow \infty$ case,
where $\phi_{r-}^- = \phi_{r+}^+$, in weak coupling 
there are two independent gaps instead of one, 
with $\phi_{r-}^- \sim \ln (1/g^2) \, \phi_{r+}^+$.

\subsection{Strong coupling}

\begin{figure} 
\hspace*{2.5cm} 
\psfig{figure=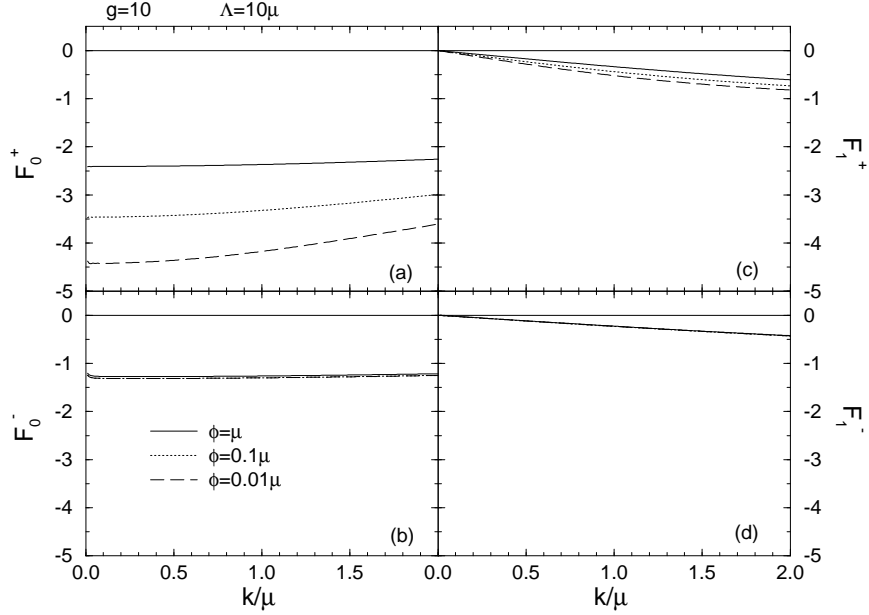,width=2.7in,height=3.5in,angle=-90}
\vspace*{-1cm}
\caption{The functions ${\cal F}_{0,1}^\pm$ as functions of
$k/\mu$ for $\phi = \mu$ (solid), $0.1\, \mu$ (dotted),
and $0.01\,\mu$ (dashed). The coupling constant is $g=10$, the
renormalization scale $\Lambda = 10\, \mu$, the temperature is taken to be
$T=0$.}
\label{fpm1}
\end{figure}

\begin{figure} 
\hspace*{2.5cm} 
\psfig{figure=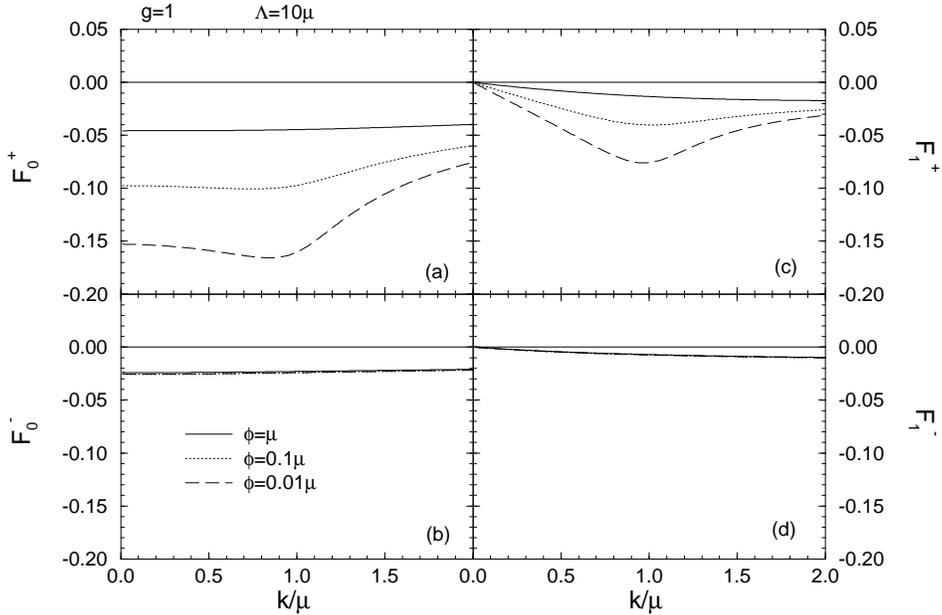,width=2.7in,height=3.5in,angle=-90}
\vspace*{-1cm}
\caption{Same as in Fig.\ \ref{fpm1}, but for $g=1$.}
\label{fpm2}
\end{figure}

In strong coupling, fermions of all energies $|{\bf q}|$ contribute 
to the functions ${\cal F}_{0,1}^{\pm}$, so that they have to be computed 
numerically. Since these functions depend on $k \equiv |{\bf k}|$, 
the set of algebraic equations (\ref{algebraicgapeqs}) 
is consistent only if this $k$ dependence is negligible.
Before actually solving (\ref{algebraicgapeqs}),
we therefore investigate the functional dependence of
${\cal F}_{0,1}^{\pm}$ on $k$ in detail.
This dependence is shown in Fig.\ \ref{fpm1} for $g=10$ and
in Fig.\ \ref{fpm2} for $g=1$ for various values of $\phi$.
The renormalization scale is $\Lambda = 10\, \mu$ in both cases.

\vspace*{1cm}

\vspace*{1cm}

As the coupling $g$ or $\phi$ become larger, the functions
${\cal F}_{0,1}^\pm$ depend less strongly on $k$. 
In weak coupling and for small $\phi$,
${\cal F}_{0,1}^-$ remain approximately constant, although they are
then small in magnitude, while ${\cal F}_{0,1}^+$ 
become peaked around the Fermi surface. In strong coupling, 
${\cal F}_1^\pm$ is smaller than ${\cal F}_0^\pm$, since
then $M_s \sim g \mu$ becomes large and the theory approaches the
NJL limit, where ${\cal F}_1^\pm \equiv 0$.
Note that changes in the renormalization scale shift the values of the
functions ${\cal F}_0^{\pm}$ by a constant amount;
a larger $\Lambda$ increases the absolute values of these functions.

Lastly we find that while the functions ${\cal F}_{0,1}^+$ change with
$\phi$, the functions ${\cal F}_{0,1}^-$  are 
nearly independent of $\phi$. The latter is easily understood for
small $|\phi| \ll \mu$, since $\phi$ only enters the integrals through 
$\epsilon^- = \sqrt{(q+\mu)^2 +\phi^2} \simeq 2\mu$ for $|\phi| \ll \mu$.
It is interesting, however, that this behavior persists also for
$|\phi|$ of order $\mu$.

\begin{figure} 
\hspace*{2.5cm} 
\psfig{figure=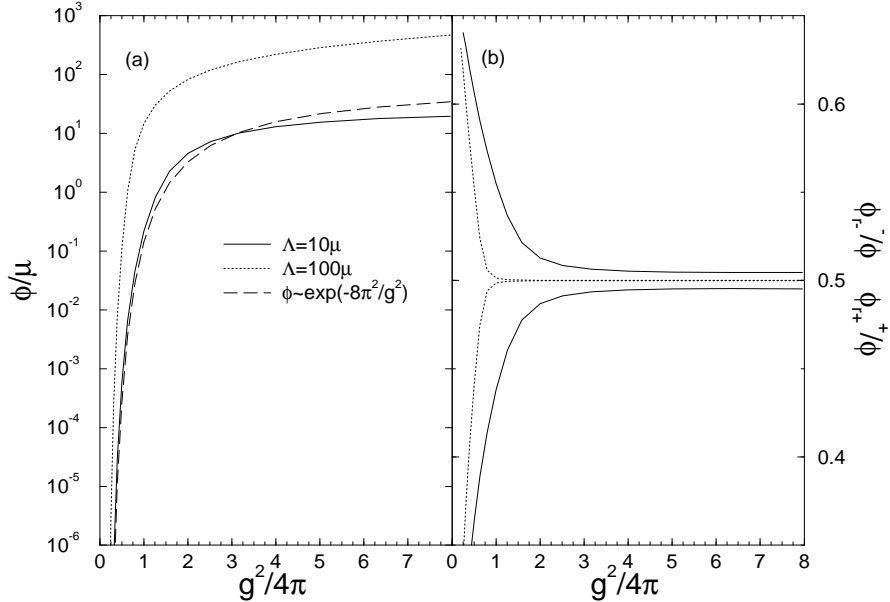,width=2.7in,height=3.5in,angle=-90}
\vspace*{-1cm}
\caption{(a) The dependence of $\phi$ (in units of $\mu$)
on $g^2/4 \pi$ for $\Lambda = 10\, \mu$ (solid) and
$100\, \mu$ (dotted). (b) The corresponding values
of $\phi_{r-}^-$ (upper set of curves) and $\phi_{r+}^+$ 
(lower set of curves), in units of $\phi$. The dashed
curve in (a) corresponds to the weak-coupling limit for $\phi$.}
\label{GapT0g}
\end{figure}

In conclusion, neglecting a possible $k$ dependence does not
appear to be a bad approximation unless the coupling becomes
substantially smaller than 1. The results of subsection \ref{wc}
are then sufficient to obtain the exponential dependence of the gaps
on the coupling; to determine the prefactor $\delta$ 
requires more sophisticated methods to 
solve the gap equations, for instance the approach of
Ref.\ \cite{khodel}. However, the gap energies
become rather small for $g < 1$, cf.\ Fig.\ \ref{GapT0g}.
We shall therefore numerically evaluate the gap equations 
(\ref{algebraicgapeqs}) only for $g \geq 1$, where the $k$ dependence can 
be safely neglected. In what follows, we then always take the values of the 
functions ${\cal F}_{0,1}^\pm$ at momentum $k=\mu$. 

Solving the gap equations (\ref{algebraicgapeqs}) numerically,
we find that the solutions satisfy
the relations $\phi_{r+}^+ = - \phi_{\ell -}^+,\,
\phi_{r-}^- = - \phi_{\ell +}^+$, as is true in the limits of
$M_s \rightarrow \infty$ and weak coupling.
According to eq.\ (\ref{0+}), this implies that
for massless fermions interacting via scalar boson exchange, 
condensation is possible only in the $0^+$ channel.

In Fig.\ \ref{GapT0g} (a) we show the dependence of the quantity
$\phi \equiv \sqrt{ \left[ \phi_{r+}^+ \right]^2 +
\left[ \phi_{\ell -}^+ \right]^2 + \left[ \phi_{r-}^- \right]^2
+ \left[ \phi_{\ell +}^- \right]^2}$ at $T=0$ on
$g^2/4 \pi$, and in Fig.\ 
\ref{GapT0g} (b) the associated behavior of $\phi_{r-}^-$ and
$\phi_{r+}^+$, normalized to $\phi$, both for different values of the
renormalization scale $\Lambda$.

\vspace*{0.5cm}

The value of $\phi$ decreases rapidly with the coupling, but values of
$\phi$ of the order of $\mu$ and larger are possible if the coupling is
sufficiently large. The value of $\phi$ increases with increasing
renormalization scale $\Lambda$. 
The overall behavior is in good agreement with the weak-coupling limit
\begin{equation}
\phi = 2\,\delta\, 
\left\{2+2\left( \ln \left[\frac{4\mu^2}{M_s^2}\right]-1 \right)^2 
\right\}^{1/2}\, \exp\left[-\frac{8\pi^2}{g^2}\right] \,\, ,
\end{equation}
which is shown by the dashed line. Here, $\delta \simeq 25\, \mu$ was taken
to fit the numerical results for $\Lambda=10\, \mu$. For a
respectively larger value, the weak-coupling limit can also
approximately fit the results for $\Lambda=100\, \mu$.

Turning to Fig.\ \ref{GapT0g} (b) we observe that,
for increasing coupling, the difference between $\phi_{r-}^-$ and
$\phi_{r+}^+$ decreases. This is in accord with our expectations, since
the mass increases $\sim g$, and consequently, the boson propagator
approaches the form
$D(p,M_s) \simeq 1/M_s^2$. We have already discussed at
the end of section \ref{II} that in this case $\phi_{r+}^+ = \phi_{r-}^-$
and $\phi_{\ell -}^+ = \phi_{\ell+}^-$.
An interesting observation is that the values for the quasi-antiparticle
gap $\phi_{r-}^-$ are larger than those for
the quasiparticle gap $\phi_{r+}^+$ for all values of $g$, not just
in weak coupling, cf.\ eq.\ (\ref{weakcoupresults}).

\subsection{Temperature dependence of the gap}

Finally, we discuss the temperature dependence
of the gap functions. As in ordinary superconductors,
the value of $\phi$ decreases with temperature, and vanishes
at $T_c$, the critical temperature for the onset of superconductivity
(or, in our case, superfluidity).
$T_c$ can be estimated in weak coupling, where the integrals
peak in a narrow region around $\mu$.
With $\phi \equiv 0$ one derives
\begin{mathletters}
\begin{eqnarray}
{\cal F}_0^+ & \simeq & - \frac{g^2}{16\pi^2}\, \,
              \ln\frac{4 \mu^2}{M_s^2} \, 
              \ln \frac{\zeta\delta}{T_c}
 \,\, , \\
{\cal F}_1^+ & \simeq & - \frac{g^2}{16\pi^2}\, 
   \left[  \ln \frac{4 \mu^2}{M_s^2} \,  - 2
   \right] \, \ln \frac{\zeta\delta}{T_c} \,\, , \\
{\cal F}_0^- & \simeq & 0\,\,\,\,  ,  \,\,\,\, {\cal F}_1^- \simeq 0 \,\,,
\end{eqnarray}
\end{mathletters}
where $\zeta \equiv 2e^\gamma/\pi$, and $\gamma$ is 
the Euler--Mascheroni constant. Using $\phi_{r+}^+ = - \phi_{\ell -}^+$,
the gap equations at $T_c$ yield the condition
\begin{equation}
1 = \frac{g^2}{8\pi^2}\, \ln \frac{\zeta\delta}{T_c} 
\end{equation}
for the critical temperature, i.e.,
\begin{equation} \label{BCS}
T_c = \frac{\zeta}{2}\, \phi_{r+}^+(T=0) \simeq 0.57 \,\phi_{r+}^+(T=0)
\,\, .
\end{equation}
This relation is identical to that found in BCS theory \cite{fetterwalecka}.

\begin{figure} 
\hspace*{2.5cm} 
\psfig{figure=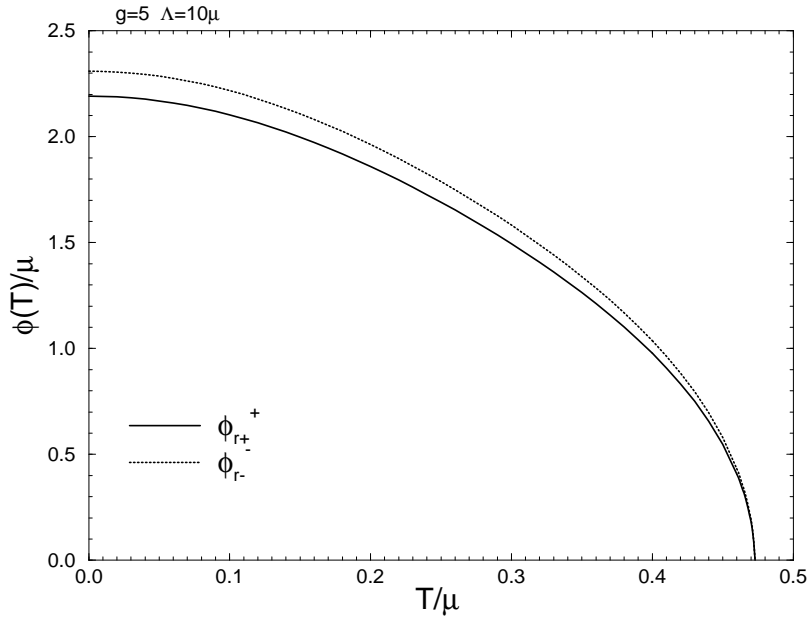,width=2.7in,height=3.5in,angle=-90}
\vspace*{-1cm}
\caption{The temperature dependence of $\phi_{r+}^+$ and $\phi_{r-}^-$ for 
$g=5,\, \Lambda = 10\, \mu$.
$T_c \simeq 0.47\, \mu$ is smaller than the value $0.57\, \phi_{r+}^+(T=0)
\simeq 1.25 \mu$ expected in weak coupling.}
\label{dofT}
\end{figure}

For arbitrary $g$, the value of $T_c$ has to be determined numerically.
For the temperature-dependent boson mass considered here, the
increase of the mass with $T$ actually leads to
smaller values for $T_c$ than expected from the BCS result (\ref{BCS}),
see Fig.\ \ref{dofT}. 
This is in agreement with 
the general expectation that larger boson masses lead to smaller
values for the gap $\phi$, and consequently to smaller values of $T_c$.
We find that $\phi_{r+}^+$ and $\phi_{r-}^-$ always vanish at the
same temperature for any value of the coupling.
\newpage
\vspace*{1cm}

\section{Conclusions}

In this paper, we have investigated superfluidity in a system
of massless fermions interacting via scalar boson exchange.
The gap matrix contains in general four independent
gap functions, corresponding to the
condensation of fermions with the same helicity and chirality.
We solved the gap equations in the mean-field approximation as
functions of the coupling $g$ and the renormalization scale $\Lambda$.
For scalar boson exchange, condensation in the $0^-$ channel does
not occur, and the number of independent gap functions reduces to
two, one for quasiparticles and one for quasi-antiparticles. The 
quasi-antiparticle gap is found to be larger than the quasiparticle gap,
by a factor $\sim \ln (1/g^2)$ in weak coupling. This is in contrast to
NJL-type models, where the point-like four-fermion interactions
do not allow for pairing with $L=S=1$ and force the equality of
the quasiparticle and quasi-antiparticle gaps.
We also analyzed the temperature dependence of the gap functions, and
found that a temperature-dependent boson mass can significantly reduce $T_c$.

Qualitatively, our model shares the feature with QCD that the
gap is exponentially small for $g \ll 1$. As can be seen from
Fig.\ \ref{GapT0g} (a), in that regime small variations of the coupling lead 
to order-of-magnitude changes in the gap. 
Moreover, the gap energies are quite sensitive to the value of the
ultraviolet cut-off $\Lambda_{\rm UV} \sim \Lambda$. Taking a small value
for the coupling,
Bailin and Love found gap energies which are extremely small, 
$\sim 10^{-3}\, \mu \sim 1$ MeV for $\mu \sim 1$ GeV \cite{bailinlove}.
In more recent studies \cite{arw,rssv,arw2,br}, much larger
gaps, $\sim 100$ MeV, were found. Such
large gaps can be obtained from the treatment of Bailin and Love simply
by taking larger coupling constants. Alternatively, the latter studies
would get smaller values for the gap by reducing the coupling constant,
or changing the form factors employed there to remove the ultraviolet
divergence of the gap integrals.

Clearly, it is essential to correctly compute the magnitude of the gap in QCD.
Studies along this line are in progress \cite{son,QCDrdpdhr}.
\\ ~~ \\
{\bf Acknowledgments}
\\ ~~ \\
We acknowledge discussions with M.\ Alford, A.\ Blaer, D.\ Blaschke,
V.J.\ Emery, M.\ Gyulassy, M.\ Laine, C.\ Manuel, 
B.\ M\"uller, V.N.\ Muthukumar,
K.\ Rajagopal, T.\ Sch\"afer, and D.T.\ Son.
D.H.R.\ thanks Columbia University's Nuclear Theory Group for
continuing access to their computing facilities.

\appendix

\section{The Mean-Field Approximation}
\label{A}

\begin{figure}
\begin{center}
\epsfxsize=4cm
\epsfysize=4cm
\leavevmode
\hbox{ \epsffile{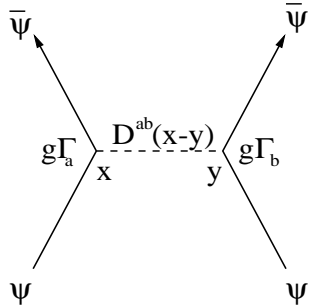}}
\end{center}
\caption{The current-current interaction.}
\label{current}
\end{figure}

Consider a statistical mechanical system at finite temperature $T$ and
chemical potential $\mu$ where fermions interact via exchange of 
an $N$-component bosonic field. 
For $N=1$ the Lagrangian is identical with eq.\ (\ref{L}).
The grand partition function of this system
is
\begin{mathletters}
\begin{eqnarray}
{\cal Z} & = & {\cal N} 
\int {\cal D} \bar{\psi}\, {\cal D}\psi\, {\cal D} \phi \, 
\exp\left\{ I[\bar{\psi},\psi,\phi]\right\} \,\, , \\
I[\bar{\psi},\psi,\phi] 
&  =  & \int_{x,y} \left(  \bar{\psi}(x) \,
\left[G_0^{+}\right]^{-1}\!(x,y)\, \psi(y)  - \frac{1}{2} \sum_{a,b=1}^N 
\phi^a(x) \, D_{ab}^{-1}(x,y)\, \phi^b(y) \right)   \nonumber \\
&  -  &  \int_x \sum_{a=1}^N\, g\, 
\bar{\psi}(x) \, \Gamma_a \, \psi(x)\, \phi^a(x)  \,\, .
\end{eqnarray}
\end{mathletters}
Here, 
\begin{equation}
\left[G_0^{\pm}\right]^{-1}\!(x,y) \equiv -i \left[ \, i \gamma \cdot \partial
\pm \mu \, \gamma_0 -m \right] \, \delta^{(4)}(x-y) \,\, .
\end{equation}
$D_{ab}(x,y)$ is the boson propagator, the structure of which need not
be further specified at this point. 
The above form includes scalar 
interactions for $N=1$, $\Gamma_1 = 1$, vector interactions for $N=4$,
$\phi^a = g^{a-1,\mu} \phi_\mu$, $\Gamma_a = g_{a-1,\mu} \gamma^\mu$, 
$D^{-1}_{ab} = g_{a-1,\mu}  [D^{-1}]^{\mu \nu} g_{b-1,\nu}$, $\mu, \nu =
0,\ldots,3$, and similarly for other interactions.
The bosonic fields can be formally integrated out with the result 
\begin{mathletters}
\begin{eqnarray}
{\cal Z} & = & {\cal N}' \left({\rm det}\, D^{-1} \right)^{-1/2}
\int {\cal D} \bar{\psi}\, {\cal D}\psi\, \exp \left\{  I[\bar{\psi},\psi]
\right\} \,\, ,\label{Z}\\
I[\bar{\psi},\psi] & = &  \int_{x,y} \left(
 \bar{\psi}(x)  \left[G_0^{+}\right]^{-1}\!(x,y) \, \psi(y)
+ \frac{g^2}{2} \sum_{a,b} \bar{\psi}(x) \, \Gamma_a \, \psi(x) \, 
D^{ab}(x,y)\, \bar{\psi}(y)\, \Gamma_b\, \psi(y)\right) \,\, .
\label{I}
\end{eqnarray}
\end{mathletters}
The last term physically corresponds
to the current-current interaction displayed in Fig.\ \ref{current}.
Since it is biquadratic in the fermion fields, the integration over
$\bar{\psi},\, \psi$ cannot be carried out. In this Appendix, we discuss
mean-field approximations to (\ref{Z}). The idea is to approximate
the biquadratic term in (\ref{I}) by a bilinear term times a {\em fermion
condensate}, which then allows for integrating over $\bar{\psi},\, \psi$.
In principle, one can either have an antifermion-fermion condensate
or a fermion-fermion condensate, the latter being the case of interest 
in describing the phenomenon of superconductivity.

\subsection{The mean-field approximation for an antifermion-fermion
condensate} \label{A1}

\begin{figure}
\begin{center}
\epsfxsize=10cm
\epsfysize=5cm
\leavevmode
\hbox{ \epsffile{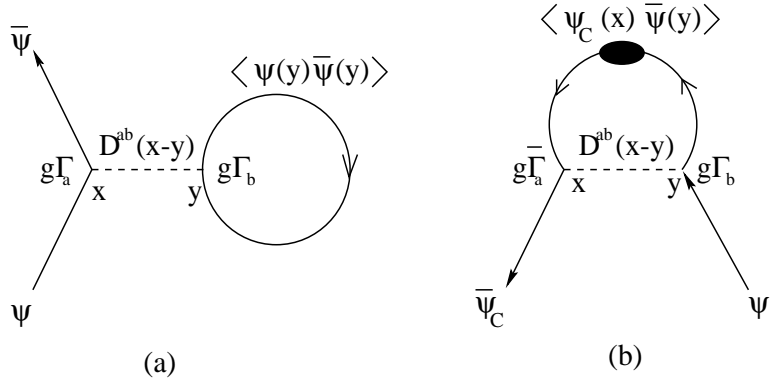}}
\end{center}
\caption{(a) The mean-field approximation for an antifermion-fermion
condensate, obtained
by contracting $\bar{\psi}(y)$ and $\psi(y)$. (b) The mean-field
approximation for a fermion-fermion condensate,
contracting $\bar{\psi}(x)$ and $\bar{\psi}(y)$.}
\label{meanfield}
\end{figure}

In the mean-field approximation one approximates two of the four fermion
fields in the last term in (\ref{I}) by their expectation value in
order to obtain a bilinear form in the fermion fields which allows
for integration over $\bar{\psi},\, \psi$ in (\ref{Z}). 
One possibility is to contract $\bar{\psi}$ and $\psi$, as shown
in Fig.\ \ref{meanfield} (a). 
The expectation value $\langle \bar{\psi} \, \psi \rangle$
corresponds to an antifermion-fermion condensate.

More specifically \cite{dhrwg}, define
\begin{equation}
j_a(x) \equiv g \, \bar{\psi}(x) \, \Gamma_a \psi(x) \,\, ,
\end{equation}
and introduce the fluctuation $\rho_a(x)$ of $j_a(x)$ around its 
expectation value $\langle j_a(x) \rangle$,
\begin{equation}
\rho_a(x) = j_a(x) - \langle j_a(x) \rangle\,\,.
\end{equation}
Then, assuming $D_{ab}(x,y) = D_{ba}(y,x)$, 
one has to first order in $\rho_a$:
\begin{eqnarray}
\lefteqn{\frac{g^2}{2} \sum_{a,b} \bar{\psi}(x) \, \Gamma_a \, \psi(x) \, 
D^{ab}(x,y)\, \bar{\psi}(y)\, \Gamma_b\, \psi(y)  \equiv \frac{1}{2}
\sum_{a,b} j_a(x) \, D^{ab}(x,y)\, j_b(y) } \nonumber \\
& \simeq & \frac{1}{2} \sum_{a,b}
\left[ \langle j_a(x) \rangle \, D^{ab}(x,y)\, \langle j_b(y) \rangle 
+ 2\, \rho_a(x) \, D^{ab}(x,y)\, \langle j_b(y) \rangle \right] \nonumber \\
& = & \frac{1}{2}
\sum_{a,b} \left[ - \langle j_a(x) \rangle \, D^{ab}(x,y)\, \langle j_b(y) 
\rangle + 2 \, j_a(x) \, D^{ab}(x,y)\, \langle j_b(y) \rangle \right] \,\, .
\end{eqnarray}
Inserting this back into (\ref{I}) and integrating over the fermion fields
in (\ref{Z}) gives the partition function in the mean-field approximation
for an antifermion-fermion condensate
\begin{equation}
{\cal Z}_{\rm \langle \bar{F}F \rangle} = {\cal N}' \left({\rm det}\, D^{-1} 
\right)^{-1/2}\, {\rm det}\, \left[G_{\rm \langle \bar{F}F \rangle}^{+}
\right]^{-1} 
\exp \left\{ - \frac{g^2}{2} \int_{x,y} \sum_{a,b} \langle \bar{\psi}(x)\,
\Gamma_a\, \psi(x) \rangle \, D^{ab}(x,y)\, \langle \bar{\psi}(y)
\, \Gamma_b \, \psi(y) \rangle \right\}  \,\, ,
\end{equation}
where the (inverse) fermion propagator is
\begin{equation}
\left[G_{\rm \langle \bar{F}F \rangle }^{+}\right]^{-1}\!(x,y) \equiv 
\left[G_0^{+}\right]^{-1}\!(x,y)
-i\, g^2 \, \delta^{(4)}(x-y)
\int_z \sum_{a,b} \Gamma_a \, D^{ab}(x,z)\, \langle \bar{\psi}(z)\,
\Gamma_b\, \psi(z) \rangle \,\, .
\end{equation}
This last equation is obviously a self-consistency equation (or gap equation,
or Schwinger-Dyson equation) for the mean-field propagator, since
\begin{equation}
\langle \bar{\psi}(z)\, \Gamma_b\, \psi(z) \rangle \equiv  {\rm Tr}
\left\{ G_{\rm \langle \bar{F}F \rangle}^{+}(z,z) \, \Gamma_b \right\} \,\, .
\end{equation}
The explicit evaluation of the partition function is facilitated assuming
translational invariance which allows to Fourier-transform all quantities
to momentum space, for details, see \cite{dhrwg}.

\subsection{The mean-field approximation for a fermion-fermion condensate}
\label{A2}

In the previous subsection \ref{A1}, we have discussed the mean-field
approximation for an antifermion-fermion condensate. Now
we discuss condensation of fermion pairs.
Instead of contracting $\bar{\psi}$ and $\psi$, we now contract
$\bar{\psi}$ and $\bar{\psi}$ (or equivalently, 
$\psi$ and $\psi$), cf.\ Fig.\ \ref{meanfield} (b). 
This leads to a fermion-fermion condensate.
More specifically, introduce the charge-conjugate spinor $\psi_C$ via
(\ref{conjugatespinor}),
and define $\bar{\Gamma}_a$ as 
\begin{equation}
\bar{\Gamma}_a = C\, \Gamma_a^T \, C^{-1}\,\, .
\end{equation} 
Then, the four-fermion term in (\ref{I}) is written as
\begin{eqnarray}
\bar{\psi}(x) \, \Gamma_a \, \psi(x)  
\, \bar{\psi}(y)\, \Gamma_b\, \psi(y) 
& = & \frac{1}{2} \left[ \bar{\psi}_C(x) \, \bar{\Gamma}_a \, \psi_C(x) \, 
\bar{\psi}(y)\, \Gamma_b\, \psi(y)  + {\rm h.c.} \right] \nonumber \\
& = & - \frac{1}{2} {\rm Tr} \left[ \gamma_0\, J^\dagger (y,x)\, \gamma_0 \,
\bar{\Gamma}_a \, J(x,y)\, \Gamma_b +{\rm h.c.} \right]
\,\, , \label{4fermion}
\end{eqnarray}
where we introduced the $4\times 4$ matrix
\begin{equation}
J_{\alpha \beta}(x,y) \equiv \psi_{C\alpha}(x) \bar{\psi}_{\beta}(y)\,\,\, ,
\,\,\,\,
J^\dagger_{\beta \alpha}(y,x) = \left[J_{\alpha \beta}(x,y) \right]^\dagger
= \left[ \gamma_0\, \psi(y)\right]_\beta \,
\left[\bar{\psi}_C(x)\, \gamma_0 \right]_{\alpha}
\,\, .
\end{equation}
Introducing the fluctuation of this matrix around its expectation value,
\begin{equation}
\rho(x,y) \equiv J(x,y) - \langle J(x,y) \rangle \,\, ,
\end{equation}
and expanding (\ref{4fermion}) to linear order in $\rho$ yields
\begin{eqnarray}
\lefteqn{\bar{\psi}(x) \, \Gamma_a \, \psi(x)  
\, \bar{\psi}(y)\, \Gamma_b\, \psi(y) }\nonumber \\
& \simeq & 
\frac{1}{2} \,{\rm Tr}\, \left[ \gamma_0\, 
\langle J^\dagger (y,x) \rangle \, \gamma_0 \,
\bar{\Gamma}_a \, \langle J(x,y) \rangle\, \Gamma_b
 -  2\, \gamma_0\, J^\dagger(y,x)\, \gamma_0 \, \bar{\Gamma}_a
\, \langle J(x,y) \rangle \, \Gamma_b +{\rm h.c.} \right] \,\, .
\end{eqnarray}
The result for the partition function
in the mean-field approximation for a fermion-fermion condensate reads
\begin{eqnarray}
{\cal Z}_{\rm \langle FF \rangle } & = & 
{\cal N}' \left( {\rm det}\, D^{-1} \right)^{-1/2} 
\exp \left\{ \frac{g^2}{4} \int_{x,y} \sum_{a,b} 
{\rm Tr}\, \left[\gamma_0 \,\langle J^\dagger (y,x) \rangle \, \gamma_0 \,
\bar{\Gamma}_a \, \langle J(x,y) \rangle\, \Gamma_b +{\rm h.c.}
\right]\, D^{ab}(x,y) \right\}  \nonumber \\
& \times &  \int {\cal D}\bar{\psi}\, {\cal D} \psi\, \exp 
\left\{  I[\bar{\psi},\psi] \right\} \,\, , \label{A16}
\end{eqnarray}
with
\begin{equation} \label{action}
I [\bar{\psi},\psi] = 
\int_{x,y} \, \left\{ \bar{\psi}(x)\, 
\left[G_0^{+}\right]^{-1}\!(x,y)\, \psi(y) + 
\frac{1}{2}  \left[ \bar{\psi}_C(x) \,
\Delta^{+}(x,y) \, \psi(y) + {\rm h.c.} \right]\right\} \,\, , 
\label{action2}
\end{equation}
where
\begin{equation} \label{gapequation2}
\Delta^{+}(x,y) \equiv 
g^2 \sum_{a,b} \bar{\Gamma}_a \, \langle \psi_C(x) \, 
\bar{\psi}(y)\rangle \, \Gamma_b \, D^{ab}(x,y)\,\, .
\end{equation}
This equation uniquely defines the gap equation in the 
mean-field approximation.
In order to solve this equation, one has to compute the expectation value
$\langle \psi_C(x) \, \bar{\psi}(y)\rangle$. This is done as follows.
The hermitean conjugate of the first term in square brackets in
(\ref{action2}) is
\begin{equation}
\left[ \bar{\psi}_C(x)\, \Delta^{+}(x,y) \, \psi(y) \right]^\dagger = 
\bar{\psi}(y) \, \Delta^{-}(y,x) \, \psi_C(x) \,\, ,
\end{equation}
where 
\begin{equation} \label{deltaminus}
\Delta^{-}(y,x) \equiv \gamma_0 \left[\Delta^{+}(x,y)\right]^\dagger 
\gamma_0\,\, .
\end{equation} 
The first term in (\ref{action2}) can be rewritten in terms of 
charge-conjugate spinors as
\begin{equation}
\bar{\psi}(x) \left[ G_0^{+} \right]^{-1}\!(x,y)\,  \psi(y) = 
\bar{\psi}_C(y) \left[ G_0^{-} \right]^{-1}\!(y,x) \, \psi_C(x) \,\, .
\end{equation}
Introducing the 8-component spinors
\begin{equation}
\Psi \equiv \left( \begin{array}{c} 
                    \psi \\
                    \psi_C 
                   \end{array}
            \right) \,\,\, , \,\,\,\,
\bar{\Psi} \equiv ( \bar{\psi} \, , \, \bar{\psi}_C ) \,\, ,
\end{equation}
the action (\ref{action2}) can be written in compact matrix notation as
\begin{equation}
I[\bar{\Psi},\Psi] = \frac{1}{2} \int_{x,y} \bar{\Psi}(x) \, 
{\cal S}^{-1}(x,y) \, \Psi(y) \,\, ,
\end{equation}
where
\begin{equation}
{\cal S}^{-1} = \left( \begin{array}{cc}
            \left[ G_0^{+} \right]^{-1} & \Delta^{-} \\  
            \Delta^{+} & \left[ G_0^{-} \right]^{-1} 
                \end{array}  \right) \,\, .
\end{equation}
Let us assume translational invariance for the gap matrix,
$\Delta^{\pm}(x,y) = \Delta^{\pm}(x-y)$. Then,
the Fourier transforms of the fields and 
$\left[ G_0^{\pm} \right]^{-1},\, \Delta^{\pm}$ are
\begin{mathletters} \label{FTall}
\begin{eqnarray}
\psi(x) = \frac{1}{\sqrt{V}} \sum_k e^{-i k \cdot x} \psi(k) \,\, & , &
\,\,\, \bar{\psi}(x) = \frac{1}{\sqrt{V}} \sum_k e^{i k \cdot x} \bar{\psi}(k)
\,\, , \\
\psi_C(x) = \frac{1}{\sqrt{V}} \sum_k e^{-i k \cdot x} \psi_C(k) \,\, & , &
\,\,\, \bar{\psi}_C (x) = \frac{1}{\sqrt{V}} \sum_k e^{i k \cdot x} 
\bar{\psi}_C(k) \,\, , \label{FT} \\
\left[ G_0^{\pm} \right]^{-1}\!(x) = \frac{T}{V} \sum_k 
e^{-ik \cdot x} \left[ G_0^{\pm} \right]^{-1}\!(k) \,\, & , & \,\,\,
\Delta^{\pm} (x) = \frac{T}{V} \sum_k e^{-ik \cdot x} 
\Delta^{\pm}(k) \,\, . \label{FT2}
\end{eqnarray}
\end{mathletters}
Note the choice of signs in the exponential factors
for $\psi_C$ and $\Delta^{-}$. From eq.\ (\ref{conjugatespinor}),
one would have expected the opposite sign. This choice has the consequence
\begin{equation} \label{A25}
\psi_C(k) = C\,  \bar{\psi}^T(-k) \,\,\, , \,\,\,\,
\bar{\psi}_C(k) = \psi^T(-k)\, C \,\,\, ,\,\,\,\,
\Delta^{-}(k) = \gamma^0 \left[\Delta^{+}(k)\right]^\dagger  \gamma^0
\,\, .
\end{equation}
It also ensures that the action is diagonal in momentum space:
\begin{equation} \label{action3}
I[\bar{\Psi},\Psi]  =  
\frac{1}{2} \sum_k \bar{\Psi}(k)\, \frac{{\cal S}^{-1}(k)}{T} \, \Psi(k) \,\, .
\end{equation}
In the conventional approach to superconductivity, the action is
only diagonalized after performing a Bogoliubov transformation. 
The choice of signs in
the Fourier transforms (\ref{FT},\ref{FT2}) 
for the charge-conjugate spinor and $\Delta^{-}$
avoids this additional complication.

In order to complete the calculation of the grand partition function
in the mean-field approximation, one has to perform the Grassmann integration
over the fermion fields $\bar{\psi}, \,\psi$. 
In the action (\ref{action3}), however,
the fields $\psi_C(k),\, \bar{\psi}_C(k)$ also enter; these are
not independent integration variables on account of
(\ref{conjugatespinor}). To proceed, from
(\ref{A25}) one derives
the identities 
\begin{equation} \label{transform}
\psi(-k) \equiv C\, \bar{\psi}_C^T(k)\,\,\,\, , \,\,\,\,
\bar{\psi}(-k) \equiv \psi_C^T(k)\,  C \,\, ,
\end{equation}
and rewrites the integration measure as
\begin{equation}
{\cal D} \bar{\psi} \, {\cal D} \psi \equiv \prod_k {\rm d} \bar{\psi}(k)\,
{\rm d} \psi(k)  = \prod_{k>0} {\rm d} \bar{\psi}(k)
\, {\rm d} \bar{\psi}(-k)\, {\rm d} \psi(k) 
\, {\rm d} \psi(-k) = \tilde{\cal N} 
\prod_{k>0} {\rm d} \bar{\psi}(k)
\, {\rm d} \psi_C(k)\, {\rm d} \psi(k)
\, {\rm d} \bar{\psi}_C(k)\,\, ,
\end{equation}
where $\tilde{\cal N}$ is the (irrelevant, since constant) Jacobian
from the transformation (\ref{transform}).
Moreover, one can show that
\begin{equation}
\frac{1}{2} \sum_k \bar{\Psi}(k)\, \frac{{\cal S}^{-1}(k)}{T} \, \Psi(k)
\equiv 
\sum_{k>0} \bar{\Psi}(k)\, \frac{{\cal S}^{-1}(k)}{T} \, \Psi(k)\,\, .
\end{equation}
Then,
\begin{equation}
\int {\cal D} \bar{\psi}\, {\cal D} \psi \, \exp\left\{ I\left[ \bar{\Psi},
\Psi \right] \right\} \equiv \tilde{\cal N} \,
{\rm det}_{k>0} \left[ \frac{{\cal S}^{-1}}{T}
\right] \equiv \tilde{\cal N}\, 
\left( {\rm det} \left[ \frac{{\cal S}^{-1}}{T} \right] \right)^{1/2} \,\, .
\end{equation}
The full propagator ${\cal S}(k)$ is determined from solving
$1={\cal S}^{-1}\, {\cal S}$, with the result
\begin{equation} \label{S}
{\cal S} = \left( \begin{array}{cc}
  G^{+} &   -  G_0^{+} \, \Delta^{-} \, G^{-} \\
 -  G_0^{-} \, \Delta^{+} \, G^{+} &  G^{-}
           \end{array}  \right) \,\, ,
\end{equation}
where all functions depend on the 4-momentum $k^\mu$ and where we have
introduced 
\begin{equation}
G^{\pm} \equiv \left\{ \left[ G_0^{\pm} \right]^{-1} - \Sigma^{\pm} 
\right\}^{-1} \,\,\, , \,\,\,\, \Sigma^{\pm} \equiv \Delta^{\mp}
\, G_0^{\mp}\, \Delta^{\pm} \,\, .
\end{equation}
The off-diagonal components of ${\cal S}$ fulfill the identity
\begin{equation}
G_0^{\mp} \, \Delta^{\pm} \, G^{\pm} = 
G^{\mp} \, \Delta^{\pm} \, G_0^{\pm} \,\, .
\end{equation}
This can be proven directly, or by solving $1 = {\cal S}\, {\cal S}^{-1}$.

\begin{figure}
\begin{center}
\epsfxsize=12cm
\epsfysize=4cm
\leavevmode
\hbox{ \epsffile{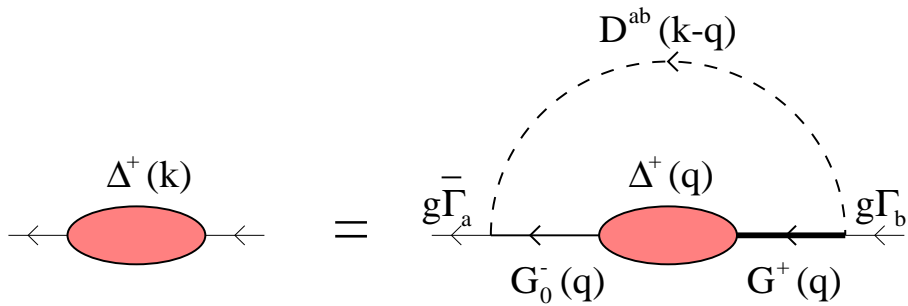}}
\end{center}
\caption{The gap equation.}
\label{gap}
\end{figure}

From eq.\ (\ref{S}), and from $\langle \Psi_\alpha(k) \bar{\Psi}_\beta(k) 
\rangle \equiv - T {\cal S}_{\alpha \beta}(k)$ (this identity is
proven e.g.\ in Appendix B of \cite{dhrwg}) one obtains
\begin{equation}
\langle \psi_C(x) \, \bar{\psi}(y)\rangle \equiv \frac{T}{V}
\sum_k e^{-ik\cdot (x-y)}\,  G_0^{-}(k) \, 
\Delta^{+}(k) \, G^{+}(k)\,\, .
\end{equation}
Inserting this into (\ref{gapequation2}) and taking the Fourier
transform, one obtains the gap equation
\begin{equation} \label{gapeq2}
\Delta^{+}(k) = g^2 \frac{T}{V} \sum_q 
\sum_{a,b} \bar{\Gamma}_a \,D^{ab}(k-q)\,  G_0^{-}(q) \, 
\Delta^{+}(q) \, G^{+}(q)\, \Gamma_b \,\, .
\end{equation}
The gap equation has a graphical representation which is derived from
Fig.\ \ref{meanfield} (b), cf.\ Fig.\ \ref{gap}.
All that changed with respect to Fig.\ \ref{meanfield} (b)
is that the explicit value of 
$\langle \psi_C(x) \, \bar{\psi}(y)\rangle$ from (\ref{S}) was used.
The blob in Fig.\ \ref{gap} stands for the gap matrix, while the
thick (thin) lines represent the full (free) propagator.

\section{The Structure of a Scalar Gap Matrix}\label{B}

In this Appendix we analyze the Dirac structure of a scalar gap matrix.
We find that in general
there are {\em eight\/} independent gap functions. Fermi statistics imposes
a powerful symmetry constraint on these functions. With respect to parity,
four of the gap functions describe condensation in the channel
of even parity, the other four in the odd-parity channel.
Only fermions with the same helicity condense. With respect to chirality,
four gap functions describe condensation of pairs with the same, the
other four that of pairs with the opposite chirality.
The chirality and helicity properties of the gap functions
suggest a different representation scheme for the gap matrix.
Within this scheme, we derive that the number of independent gap functions 
reduces to {\em four\/} in the ultrarelativistic limit. 

\subsection{Dirac structure}

The Fourier transform $\Delta (k)$ of the (translationally invariant)
scalar gap matrix $\Delta(x-y)$ can be expanded in the basis
of the 16 linearly independent $4 \times 4$ matrices. However, 
for condensation in the $J=0$ channel, all Lorentz indices have to be 
contracted with either the 4-momentum, $k^\mu$, or, because of the
presence of a medium, with the respective 4-velocity of this medium, $u^\mu$.
This leaves the eight matrices
\begin{equation} \label{basis}
{\bf 1} \, , \,\, \gamma \cdot k \, , \,\, \gamma \cdot u\, , \,\, 
\gamma \cdot k \,  \gamma \cdot u \, , \,\,
\gamma_5 \, , \,\, \gamma_5 \gamma \cdot k \, , \,\, 
\gamma_5 \, \gamma \cdot u \, , \,\,
\gamma_5 \, \gamma \cdot k \, \gamma \cdot u\,\, .
\end{equation}
In the rest frame of the medium, $u^\mu = (1,{\bf 0})$, the most
general ansatz for $\Delta$ is then
\begin{equation} \label{expansion}
\Delta = \Delta_1 \,\gamma_5 
+ \Delta_2\, \bbox{\gamma} \cdot \hat{\bf k} \gamma_0 \gamma_5 
+ \Delta_3\, \gamma_0 \gamma_5
+ \Delta_4 
+ \Delta_5 \, \bbox{\gamma} \cdot \hat{\bf k} \gamma_0 
+ \Delta_6 \, \bbox{\gamma} \cdot \hat{\bf k}
+ \Delta_7 \, \bbox{\gamma} \cdot \hat{\bf k} \gamma_5
+ \Delta_8 \, \gamma_0 \,\, ,
\end{equation}
where $\hat{\bf k} \equiv {\bf k}/|{\bf k}|$, and where
the gap matrix $\Delta$ as well as the gap functions
$\Delta_n$ depend on $k_0,\, {\bf k}$. The notation follows that
of Bailin and Love \cite{bailinlove}.

\subsection{Symmetry properties} \label{symm}

The antisymmetry of the fermion fields 
provides a powerful constraint on the gap functions $\Delta_n$.
With (\ref{conjugatespinor}) and $C = -C^{-1}$ one rewrites
\begin{equation}
\int_{x,y} \bar{\psi}_C(x) \, \Delta (x-y) \, \psi(y)
= - \int_{y,x} \psi^T(x) \, \left[ \Delta(y-x)\right]^T \, 
\bar{\psi}_C^T(y) = \int_{y,x} \bar{\psi}_C (x) \, C^{-1} \,
\left[ \Delta(y-x) \right]^T \, C \, \psi(y) ,
\end{equation}
i.e., in Fourier space
\begin{equation}
C\, \Delta (k)\, C^{-1} = \Delta^T(-k) \,\, .
\end{equation}
Using $C\, \gamma_\mu \, C^{-1} = - \gamma_\mu^T$, this implies:
\begin{equation}
\Delta_n(k) = + \Delta^T_n(-k)\,\,\, , \,\,\, n=1,\ldots,6\,\,\,\, , \,\,\,\,
\Delta_n(k) = - \Delta^T_n(-k)\,\,\, , \,\,\, n=7,8 \,\, .
\end{equation}
If we neglect the energy-momentum dependence of the gap functions
$\Delta_n$, as we do in our solution of the gap equations,  
then this equation demands that $\Delta_1 , \ldots, \, \Delta_6$ are
{\em symmetric\/} matrices in the space of internal degrees of freedom, 
while $\Delta_7$ and $\Delta_8$ are {\em antisymmetric}. 
In QCD, for example, $\Delta^{ij}_{n,fg} = \Delta^{ji}_{n,gf},\, 
n=1,\ldots, \, 6$, while $\Delta^{ij}_{n,fg} = - \Delta^{ji}_{n,gf},\,n=7,8$
[for notation see the discussion preceding eq.\ (\ref{L})].
If the $\Delta_n$ do not have internal degrees of freedom, then
$\Delta_7$ and $\Delta_8$ must vanish, as noted by
Bailin and Love \cite{bailinlove}. We stress, however, that 
$\Delta_7$ and $\Delta_8$ need not vanish when they carry
internal degrees of freedom.

\subsection{Parity} \label{parity}

Under a parity transformation, $t \rightarrow t,\, {\bf x} 
\rightarrow - {\bf x}$, so 3-vectors change their sign.
On Dirac spinors, a parity transformation is effected by \cite{gross} 
$S(P) = \eta_P \gamma_0$, where $\eta_P = \pm 1$ is
the intrinsic parity of a particle. Thus $\gamma_0$ has even parity,
$S^{-1}(P)\, \gamma_0\, S(P) = \gamma_0$, while $\gamma^i$ has odd parity,
$S^{-1}(P)\, \gamma^i \, S(P) = - \gamma^i$.
Wave functions transform as $\psi'(t,{\bf x}) = S(P)\, 
\psi(t, -{\bf x}) = \eta_P \gamma_0 \,
\psi(t,-{\bf x})$, i.e., with (\ref{conjugatespinor}),
$\bar{\psi}_C'(t,{\bf x}) = - \bar{\psi}_C(t,-{\bf x})\, S(P)$, which
shows that the spinor $\bar{\psi}_C$ has opposite parity 
from $\psi$. 

We rewrite the action (\ref{effaction}) in Fourier space, with the
conventions listed in eq.\ (\ref{FTall}):
\begin{equation} \label{effaction2}
I[\bar{\psi},\psi] = \sum_k \left\{
\bar{\psi}(k)\, \left[G_0^{+}\right]^{-1}\!(k) 
\, \psi(k) + \frac{1}{2} \left[ \bar{\psi}_C(k) \, \Delta(k) \, \psi(k)
+ {\rm h.c.}\right] \right\} \,\, .
\end{equation}
Then, the term
\begin{equation}
\bar{\psi}_C(k) \left(\Delta_1 \,\gamma_5 + \Delta_2\, \bbox{\gamma}
\cdot \hat{\bf k} \gamma_0 \gamma_5 + \Delta_3\, \gamma_0 \gamma_5
+ \Delta_7 \, \bbox{\gamma} \cdot \hat{\bf k} \gamma_5 \right) \, \psi(k)
\end{equation}
represents condensation in the {\em even-parity channel}, while the term
\begin{equation} 
\bar{\psi}_C(k) \left(\Delta_4 + \Delta_5 \, \bbox{\gamma} \cdot 
\hat{\bf k} \gamma_0 + \Delta_6 \, \bbox{\gamma} \cdot \hat{\bf k} 
+ \Delta _8\, \gamma_0 \right) \, \psi(k)
\end{equation}
represents condensation in the {\em odd-parity\/} channel.

\subsection{Helicity}

The spinors $\psi,\, \bar{\psi},\, \psi_C,\, \bar{\psi}_C$ in the
effective action (\ref{effaction2})
can be decomposed with respect to their helicity. 
The helicity projector is given by 
\begin{equation} \label{heliproj}
{\cal P}_{\pm}({\bf k}) \equiv \frac{1 \pm \gamma_5 \gamma_0 \bbox{\gamma}
\cdot {\bf k}}{2}\,\, ,
\end{equation}
and we denote the helicity-projected spinors by
\begin{mathletters}
\begin{eqnarray}
\psi_{\pm}(k) =  {\cal P}_{\pm}({\bf k}) \, \psi(k)\,\,\, & , &
\,\,\,\,
\bar{\psi}_{\pm}(k) =  \bar{\psi}(k)\, {\cal P}_{\pm}({\bf k}) \,\, , \\
\psi_{C \pm}(k) =  {\cal P}_{\pm}({\bf k})\, \psi_C(k)\,\,\, & , &
\,\,\,\,
\bar{\psi}_{C \pm}(k) =  \bar{\psi}_C(k)\, {\cal P}_{\pm}({\bf k})\,\, .
\end{eqnarray}
\end{mathletters}
The second equation results from the fact that due to our sign convention
in (\ref{FT}), in Fourier space $\bar{\psi}_C(k) = \psi^T(-k)\, C$.

The inverse free fermion propagator as well as the gap 
matrix (\ref{expansion}) commute with the helicity projector,
\begin{equation}
\left[ \left[G_0^{+} \right]^{-1}\!(k)\,,\, {\cal P}_{\pm}({\bf k}) \right]
= \left[\Delta(k) \,, \, {\cal P}_{\pm}({\bf k}) \right] = 0 \,\, .
\end{equation}
This means that the action (\ref{effaction2}) does not mix
states of different helicity,
\begin{equation}
I [\bar{\psi}, \psi] \equiv \sum_{s=\pm} I
[\bar{\psi}_s ,\psi_s ]\,\, .
\end{equation}
As a consequence, condensation in a scalar ($J=0$) channel
{\em can only occur between fermions of the same helicity\/} ($++$ or $--$).
For a physical explanation, see the discussion following eq.\ (\ref{10}).

\subsection{Chirality} \label{chiral}

The chirality projector is given by 
\begin{equation} \label{chiralproj}
{\cal P}_{r,\ell} \equiv \frac{1 \pm \gamma_5}{2} \,\, .
\end{equation}
Let us introduce right- and left-handed spinors via
\begin{mathletters}
\begin{eqnarray}
\psi_r \equiv {\cal P}_r \, \psi \,\,\,\,  & , &\,\,\,\,
\psi_{\ell} \equiv {\cal P}_{\ell} \, \psi \,\, ,  \\
\bar{\psi}_r \equiv \bar{\psi} \, {\cal P}_{\ell} \,\,\,\, & , & \,\,\,\,
\bar{\psi}_{\ell} \equiv \bar{\psi} \, {\cal P}_r \,\, .
\end{eqnarray}
\end{mathletters}
With (\ref{conjugatespinor}), one then derives
\begin{mathletters}
\begin{eqnarray}
\psi_{C r} \equiv {\cal P}_{\ell} \, \psi_C \,\,\,\, & , & \,\,\,\,
\psi_{C \ell} \equiv {\cal P}_r \, \psi_C \,\, ,  \\
\bar{\psi}_{C r} \equiv \bar{\psi}_C \, {\cal P}_r \,\,\,\, & , &
\,\,\,\,
\bar{\psi}_{C \ell} \equiv \bar{\psi}_C \, {\cal P}_{\ell} \,\, .
\end{eqnarray}
\end{mathletters}
Let us now investigate the effect of the chirality projectors
on the gap matrix (\ref{expansion}). From 
$\gamma^\mu \gamma_5 = - \gamma_5 \gamma^\mu$ and
${\cal P}_{r,\ell}^2 = {\cal P}_{r, \ell},\, {\cal P}_r {\cal P}_{\ell} =0$
one derives
\begin{equation}
{\cal P}_{r, \ell} \, \Delta \, {\cal P}_{r, \ell}
= \Delta_1 \, \gamma_5 + \Delta_2\, \bbox{\gamma}
\cdot \hat{\bf k} \gamma_0 \gamma_5 + \Delta_4 
+ \Delta_5 \, \bbox{\gamma} \cdot \hat{\bf k} \gamma_0 \,\, ,
\end{equation}
while
\begin{equation}
{\cal P}_{\ell,r} \, \Delta \, {\cal P}_{r,\ell}
= \Delta_3\, \gamma_0 \gamma_5 
+ \Delta_6 \, \bbox{\gamma} \cdot \hat{\bf k} 
+ \Delta_7 \, \bbox{\gamma} \cdot \hat{\bf k} \gamma_5
+ \Delta_8 \, \gamma_0  \,\, ,
\end{equation}
This result means that 
$\Delta_1,\, \Delta_2, \, \Delta_4$, and $\Delta_5$ 
are gap functions describing condensation of fermion
pairs with the {\em same\/} chirality (right-right or left-left), while
$\Delta_3,\, \Delta_6,\, \Delta_7,$ and $\Delta_8$ 
are gap functions describing condensation of
fermion pairs with the {\em opposite\/} chirality (right-left and
left-right). 

Let us summarize the results of this and the preceding subsections.
Only particles with the same helicity can condense to form a
scalar ($J=0$) condensate.
In the general ansatz (\ref{expansion}), the term
\begin{equation} \label{evenparity}
\Delta_1 \, \gamma_5 + \Delta_2\, \bbox{\gamma}
\cdot \hat{\bf k} \gamma_0 \gamma_5
\end{equation}
describes the condensation of 
fermions with 
the {\em same\/} chirality in the {\em even-parity\/} channel, while 
\begin{equation} \label{oddparity}
\Delta_4 + \Delta_5 \, \bbox{\gamma} \cdot \hat{\bf k} \gamma_0 
\end{equation}
describes condensation of fermions with 
the same chirality 
in the {\em odd-parity\/} channel, and
\begin{equation}
\Delta_3 \,\gamma_0 \gamma_5  + \Delta_7 \,
\bbox{\gamma} \cdot \hat{\bf k} \gamma_5
\end{equation}
describes condensation of 
{\em opposite-chirality\/} fermions
in the {\em even-parity\/} channel, while
\begin{equation} 
\Delta_6 \,\bbox{\gamma} \cdot \hat{\bf k}
+ \Delta_8 \, \gamma_0 
\end{equation}
describes condensation of opposite-chirality fermions in the 
{\em odd-parity\/} channel.

\subsection{Quasiprojector representation of the gap matrix}

The properties of the gap matrix with respect to chirality and
helicity suggest a somewhat different representation. Instead of the
eight matrices (\ref{basis}),
one can use the eight matrices constructed
from chirality, helicity, and energy projectors.
We use the following projectors onto states of positive and negative energy for
free particles:
\begin{equation} \label{energyproj}
\Lambda^{\pm} ({\bf k}) 
\equiv \frac{E_{\bf k} \pm (\gamma_0 \bbox{\gamma} \cdot {\bf k}
+ m \gamma_0)}{2\, E_{\bf k}} \equiv \frac{1 \pm (\beta_{\bf k} \gamma_0 
\bbox{\gamma} \cdot \hat{\bf k} + \alpha_{\bf k} \gamma_0)}{2} \,\, ,
\end{equation}
where $\alpha_{\bf k} \equiv m/E_{\bf k},\, \beta_{\bf k} 
\equiv |{\bf k}|/E_{\bf k}$.
These differ by $\gamma_0$ and the
normalization of the spinors from the commonly used projectors \cite{gross},
but have the advantage that they are regular in the limit $m\rightarrow 0$.

The helicity projectors commute with either the chirality or these
energy projectors,
\begin{equation}
[ {\cal P}_{r,\ell}\, , \, {\cal P}_{\pm}({\bf k}) ] = [ {\cal P}_{\pm} 
({\bf k}) \, , \, \Lambda^{\pm}({\bf k}) ] = 0 \,\, ,
\end{equation}
but, for finite $m$, as massive spinors are not
eigenstates of chirality, the energy projectors do not commute with the
chirality projectors,
\begin{equation} \label{noncommute}
[ {\cal P}_r\, , \, \Lambda^{\pm}({\bf k}) ] = 
\mp \alpha_{\bf k} \, \gamma_0 \gamma_5
\,\,\, , \,\,\,\, 
[ {\cal P}_{\ell}\, , \, \Lambda^{\pm} ({\bf k})] = 
\pm \alpha_{\bf k} \, \gamma_0 \gamma_5
\,\, .
\end{equation}

Let us introduce the ``quasiprojectors''
\begin{equation} \label{quasiprojectors}
{\cal P}^{\pm}_{r,\ell \,\pm} ({\bf k})
\equiv {\cal P}_{r,\ell}\, {\cal P}_{\pm}({\bf k}) \, 
\Lambda^{\pm} ({\bf k}) \,\, .
\end{equation}
These eight quantities constitute a basis which is equivalent to (\ref{basis}).
Therefore, the general gap matrix (\ref{expansion}) can alternatively 
be written as
\begin{equation}
\Delta   =    \phi_{r +}^{+} {\cal P}_{r +}^{+}
            + \phi_{\ell +}^{+} {\cal P}_{\ell +}^{+}
            + \phi_{r -}^{+} {\cal P}_{r -}^{+}
            + \phi_{\ell -}^{+} {\cal P}_{\ell -}^{+}
            + \phi_{r +}^{-} {\cal P}_{r +}^{-}
            + \phi_{\ell +}^{-} {\cal P}_{\ell +}^{-}
            + \phi_{r -}^{-} {\cal P}_{r -}^{-}
            + \phi_{\ell -}^{-} {\cal P}_{\ell -}^{-} \,\, , \label{expansion2}
\end{equation}
with an obvious notation for the new gap functions $\phi_{r,\ell\, \pm}^{\pm}$.
The old gap functions $\Delta_n$ can be expressed in terms of
the new gap functions $\phi$:
\begin{mathletters} \label{DXY}
\begin{eqnarray}
\Delta_1 & = & \frac{1}{8} \left\{ \left(1 + \beta_{\bf k}\right)
\left[ \phi_{r+}^{+} - \phi_{\ell -}^{+} + \phi_{r-}^{-} - \phi_{\ell +}^{-} 
\right] + \left(1 - \beta_{\bf k}\right)
\left[ \phi_{r-}^{+} - \phi_{\ell +}^{+} + \phi_{r+}^{-} - \phi_{\ell -}^{-} 
\right] \right\}\,\, ,\\
\Delta_2 & = & \frac{1}{8} \left\{ \left(1 + \beta_{\bf k}\right)
\left[- \phi_{r+}^{+} + \phi_{\ell -}^{+} + \phi_{r-}^{-} - \phi_{\ell +}^{-} 
\right] + \left(1 - \beta_{\bf k}\right)
\left[  \phi_{r-}^{+} - \phi_{\ell +}^{+} - \phi_{r+}^{-} + \phi_{\ell -}^{-} 
\right] \right\}\,\, ,\\
\Delta_3 & = & \frac{\alpha_{\bf_k}}{8} 
\left[ - \phi_{r+}^{+} - \phi_{r-}^{+} + \phi_{\ell +}^{+} + \phi_{\ell -}^{+} 
       + \phi_{r+}^{-} + \phi_{r-}^{-} - \phi_{\ell +}^{-} - \phi_{\ell -}^{-} 
\right] \,\, ,\\
\Delta_4 & = & \frac{1}{8} \left\{ \left(1 + \beta_{\bf k}\right)
\left[ \phi_{r+}^{+} + \phi_{\ell -}^{+} + \phi_{r-}^{-} + \phi_{\ell +}^{-} 
\right] + \left(1 - \beta_{\bf k}\right)
\left[ \phi_{r-}^{+} + \phi_{\ell +}^{+} + \phi_{r+}^{-} + \phi_{\ell -}^{-} 
\right] \right\}\,\, ,\\
\Delta_5 & = & \frac{1}{8} \left\{ \left(1 + \beta_{\bf k}\right)
\left[ - \phi_{r+}^{+} - \phi_{\ell -}^{+} + \phi_{r-}^{-} + \phi_{\ell +}^{-}
\right] + \left(1 - \beta_{\bf k}\right)
\left[ \phi_{r-}^{+} + \phi_{\ell +}^{+} - \phi_{r+}^{-} - \phi_{\ell -}^{-} 
\right] \right\}\,\, ,\\
\Delta_6 & = & \frac{\alpha_{\bf_k}}{8} 
\left[ - \phi_{r+}^{+} + \phi_{r-}^{+} + \phi_{\ell +}^{+} - \phi_{\ell -}^{+} 
       + \phi_{r+}^{-} - \phi_{r-}^{-} - \phi_{\ell +}^{-} + \phi_{\ell -}^{-}
\right] \,\, ,\\
\Delta_7 & = & \frac{\alpha_{\bf_k}}{8} 
\left[   \phi_{r+}^{+} - \phi_{r-}^{+} + \phi_{\ell +}^{+} - \phi_{\ell -}^{+} 
       - \phi_{r+}^{-} + \phi_{r-}^{-} - \phi_{\ell +}^{-} + \phi_{\ell -}^{-}
\right] \,\, ,\\
\Delta_8 & = & \frac{\alpha_{\bf_k}}{8} 
\left[   \phi_{r+}^{+} + \phi_{r-}^{+} + \phi_{\ell +}^{+} + \phi_{\ell -}^{+} 
       - \phi_{r+}^{-} - \phi_{r-}^{-} - \phi_{\ell +}^{-} - \phi_{\ell -}^{-}
\right] \,\, .
\end{eqnarray}
\end{mathletters}
The new basis (\ref{quasiprojectors}) is complete,
\begin{equation}
\sum_{h=r,\ell}\sum_{s= \pm} \sum_{e=\pm} {\cal P}_{hs}^{e} ({\bf k}) 
= 1 \,\, .
\end{equation}
However, using (\ref{noncommute}) we find that 
the quasiprojectors (\ref{quasiprojectors})
are not true projectors:
\begin{mathletters}
\begin{eqnarray}
{\cal P}_{r \pm}^+\, {\cal P}_{r \pm}^+ = {\cal P}_{r \pm}^+
+ \frac{ \alpha_{\bf k}}{2} \, \gamma_0 \gamma_5 \,{\cal P}_{\ell \pm}^+
\,\,\,\, & , & \,\,\,\, 
{\cal P}_{r \pm}^+\, {\cal P}_{\ell \pm}^+  = 
- \frac{\alpha_{\bf k}}{2} \, \gamma_0 \gamma_5 \, {\cal P}_{\ell \pm}^+
\,\, , \\
{\cal P}_{\ell \pm}^+\, {\cal P}_{r \pm}^+ = 
 \frac{\alpha_{\bf k}}{2} \, \gamma_0 \gamma_5 \,{\cal P}_{r \pm}^+
\,\,\,\, & , & \,\,\,\, 
{\cal P}_{\ell \pm}^+\, {\cal P}_{\ell \pm}^+= {\cal P}_{\ell \pm}^+
- \frac{\alpha_{\bf k}}{2} \, \gamma_0 \gamma_5\, {\cal P}_{r \pm}^+ 
\,\, , \\
{\cal P}_{r \pm}^- \, {\cal P}_{r \pm}^- = {\cal P}_{r \pm}^-
- \frac{\alpha_{\bf k}}{2} \, \gamma_0 \gamma_5 \, {\cal P}_{\ell \pm}^-
\,\,\,\, & , & \,\,\,\, 
{\cal P}_{r \pm}^- \, {\cal P}_{\ell \pm}^- = 
\frac{ \alpha_{\bf k}}{2} \, \gamma_0 \gamma_5 \,{\cal P}_{\ell \pm}^- 
\,\, , \\
{\cal P}_{\ell \pm}^- \, {\cal P}_{r \pm}^-  = 
- \frac{\alpha_{\bf k}}{2} \, \gamma_0 \gamma_5 \, {\cal P}_{r \pm}^-
\,\,\,\, & , & \,\,\,\, 
{\cal P}_{\ell \pm}^- \, {\cal P}_{\ell \pm}^- = {\cal P}_{\ell \pm}^-
+ \frac{\alpha_{\bf k}}{2} \, \gamma_0 \gamma_5 \, {\cal P}_{r \pm}^- 
\,\, , \\
{\cal P}_{r \pm}^+ \, {\cal P}_{r \pm}^- = 
\frac{\alpha_{\bf k}}{2} \, \gamma_0 \gamma_5 \, {\cal P}_{\ell \pm}^-
\,\,\,\, & , & \,\,\,\, 
{\cal P}_{r \pm}^+ \, {\cal P}_{\ell \pm}^- = 
- \frac{\alpha_{\bf k}}{2} \, \gamma_0 \gamma_5 \, {\cal P}_{\ell \pm}^- 
\,\, , \\
{\cal P}_{\ell \pm}^+ \, {\cal P}_{r \pm}^-  = 
 \frac{\alpha_{\bf k}}{2} \, \gamma_0 \gamma_5 \, {\cal P}_{r \pm}^-
\,\,\,\, & , & \,\,\,\, 
{\cal P}_{\ell \pm}^+ \, {\cal P}_{\ell \pm}^-  = 
- \frac{\alpha_{\bf k}}{2} \, \gamma_0 \gamma_5 \, {\cal P}_{r \pm}^- 
\,\, , \\
{\cal P}_{r \pm}^- \, {\cal P}_{r \pm}^+  =
- \frac{ \alpha_{\bf k}}{2} \, \gamma_0 \gamma_5\, {\cal P}_{\ell \pm}^+
\,\,\,\, & , & \,\,\,\, 
{\cal P}_{r \pm}^- \, {\cal P}_{\ell \pm}^+ =  
\frac{\alpha_{\bf k}}{2} \, \gamma_0 \gamma_5\, {\cal P}_{\ell \pm}^+
\,\, , \\
{\cal P}_{\ell \pm}^- \, {\cal P}_{r \pm}^+ = 
- \frac{\alpha_{\bf k}}{2} \, \gamma_0 \gamma_5\, {\cal P}_{r \pm}^+
\,\,\,\, & , & \,\,\,\, 
{\cal P}_{\ell \pm}^-\, {\cal P}_{\ell \pm}^+ =
 \frac{\alpha_{\bf k}}{2} \, \gamma_0 \gamma_5 \,{\cal P}_{r \pm}^+ \,\, 
\end{eqnarray} \label{quasiprojprod}
\end{mathletters}
(the argument ${\bf k}$ was dropped for convenience). 

For the remainder of this subsection, we consider the 
ultrarelativistic limit, $m =\alpha_{\bf k} = 0 \, , \,
\beta_{\bf k} = 1$. In this limit, 
the quasiprojectors become {\em true\/} projectors. Moreover,
four of the eight projectors vanish,
\begin{equation} \label{zeros}
{\cal P}_{r -}^{+} = {\cal P}_{\ell +}^{+} = {\cal P}_{r +}^{-} 
= {\cal P}_{\ell -}^{-} = 0 \,\,\,\, \,\,\,\, (m=0)\,\, .
\end{equation}
This is another way of stating that there are no 
massless right- (left-) handed fermions with negative (positive)
helicity and positive energy, and no right- (left-) handed fermions with 
positive (negative) helicity and negative energy.
The reduction in the number of projectors can
be intuitively understood noting that massless fermions can be described in 
terms of 2-component Weyl spinors which require only the simpler algebra
of linearly independent $2 \times 2$ matrices.

Another consequence of the ultrarelativistic limit is that 
one of the three projectors in (\ref{quasiprojectors}) is redundant:
\begin{mathletters} \label{B31}
\begin{eqnarray}
{\cal P}_{r+,\ell-}^+ ({\bf k}) & \equiv & {\cal P}_{r, \ell} \, 
{\cal P}_{\pm} ({\bf k}) \equiv {\cal P}_{r, \ell} \, \Lambda^+ ({\bf k})
\equiv {\cal P}_\pm({\bf k}) \, \Lambda^+ ({\bf k})\,\,, \\
{\cal P}_{r-,\ell+}^- ({\bf k}) & \equiv & {\cal P}_{r, \ell} \, 
{\cal P}_{\mp} ({\bf k}) \equiv {\cal P}_{r, \ell} \, \Lambda^- ({\bf k})
\equiv {\cal P}_{\mp}({\bf k}) \, \Lambda^- ({\bf k})\,\, .
\end{eqnarray}
\end{mathletters}
Thus, we could use any two of the three projectors for chirality, helicity,
and energy to construct (\ref{quasiprojectors}). 
We keep all three indices, however, because it facilitates the 
physical interpretation of our results.

When $m=0$, as the four projectors (\ref{zeros}) vanish,
the eight independent gap functions reduce to {\em four\/}, and
eq.\ (\ref{expansion2}) becomes
\begin{equation} \label{URlimit2}
\Delta  = \phi_{r +}^{+} {\cal P}_{r +}^{+}
        + \phi_{\ell -}^{+} {\cal P}_{\ell -}^{+}
        + \phi_{r -}^{-} {\cal P}_{r -}^{-}
        + \phi_{\ell +}^{-} {\cal P}_{\ell +}^{-} \,\, .
\end{equation}
Using eq.\ (\ref{DXY}) we obtain for the gap functions $\Delta_n$:
\begin{mathletters} \label{deltaurlimit}
\begin{eqnarray}
\Delta_1 & = & \frac{1}{4} 
\left[ \phi_{r +}^{+} - \phi_{\ell -}^{+} + \phi_{r -}^{-} - 
\phi_{\ell +}^{-} \right] \,\, ,\\
\Delta_2 & = & \frac{1}{4}
\left[ - \phi_{r +}^{+} + \phi_{\ell -}^{+} + \phi_{r -}^{-} - 
\phi_{\ell +}^{-} \right] \,\, ,\\
\Delta_4 & = & \frac{1}{4}
\left[ \phi_{r +}^{+} + \phi_{\ell -}^{+} + \phi_{r -}^{-} + 
\phi_{\ell +}^{-} \right] \,\, ,\\
\Delta_5 & = & \frac{1}{4}
\left[ - \phi_{r +}^{+} - \phi_{\ell -}^{+} + \phi_{r -}^{-} + 
\phi_{\ell +}^{-} \right] \,\, ,\\
\Delta_3 & = & \Delta_6 = \Delta_7 = \Delta_8  =  0 \,\, .
\end{eqnarray}
\end{mathletters}
The use of the projectors makes it clear that
there are only four condensates for massless fermions.
This was not apparent previously \cite{bailinlove}.

Another consequence of (\ref{URlimit2}) is that the action (\ref{effaction2})
decomposes into four parts; with $\psi_{r,\ell\, \pm}^{\pm} \equiv 
{\cal P}_{r,\ell \pm}^{\pm} \psi$,
\begin{mathletters} \label{B34}
\begin{eqnarray}
I[\bar{\psi},\psi] & = & I[\bar{\psi}_{r+}^+,\psi_{r+}^+]
+ I[\bar{\psi}_{\ell -}^+,\psi_{\ell -}^+] 
+ I[\bar{\psi}_{r -}^-,\psi_{r -}^-]
+ I[\bar{\psi}_{\ell +}^-,\psi_{\ell +}^-] \,\, , \\
I[\bar{\psi}_{hs}^e,\psi_{hs}^e] & \equiv & \sum_k \left\{
\bar{\psi}_{hs}^e(k)\, \left[G_0^{+}\right]^{-1}\!(k) 
\, \psi_{hs}^e(k) + \frac{1}{2} \left[ \bar{\psi}_{C\,hs}^e(k) 
\, \Delta(k) \, \psi_{hs}^e(k)
+ {\rm h.c.}\right] \right\} \,\, ,
\end{eqnarray}
\end{mathletters}
where $h=r,\ell$, $s=\pm$, $e=\pm$.
We draw the important conclusion that
{\em in the ultrarelativistic limit, there is condensation only
of fermions with the same helicity and the same chirality.}

In the scalar model we find that $\phi_{r+}^+ = - \phi_{\ell-}^+$
and $\phi_{r-}^- = - \phi_{\ell +}^-$; thus 
\begin{equation} \label{0+}
\Delta_1 = \frac{1}{2} \left[ \phi_{r+}^+ +\phi_{r-}^- \right]
\,\,\,\, ,\,\,\,\, 
\Delta_2 = \frac{1}{2} \left[- \phi_{r+}^+ +\phi_{r-}^- \right]
\,\,\,\, , \,\,\,\,
\Delta_4 = \Delta_5 = 0 \,\, .
\end{equation}
From eqs.\ (\ref{evenparity}), (\ref{oddparity})  we conclude
that condensation occurs only in the $0^+$ channel, and not
the $0^-$ channel.

\section{Renormalizing ${\cal F}_0^\pm$} \label{C}

At $T=0$, we rewrite the function ${\cal F}_0^\pm$ as 
\begin{equation} \label{F0}
{\cal F}_0^\pm = -  \frac{g^2}{4} \int \frac{{\rm d}^3 {\bf q}}{(2 \pi)^3}
\left[ D({\bf k} - {\bf q},M_s)\, \frac{1}{\epsilon^\pm} - 
D({\bf k} - {\bf q},0)\, \frac{1}{|{\bf q}|} \right] + {\cal F}_{CT} \,\, ,
\end{equation}
with the {\em counter term}
\begin{equation} \label{CT}
{\cal F}_{CT} \equiv -  \frac{g^2}{4} \int \frac{{\rm d}^3 {\bf q}}{(2 \pi)^3}
D({\bf k} - {\bf q},0) \, \frac{1}{|{\bf q}|} \,\, .
\end{equation}
We assume that the renormalized boson mass vanishes at zero density
and temperature, and so in the counter term we take the boson to have 
zero mass as well.

The first integral in (\ref{F0}) is now ultraviolet-finite. The counter
term (\ref{CT}) can be computed via dimensional regularization. First
rewrite (\ref{CT}), using
\cite{thooft}
\begin{equation}
\frac{1}{a^\alpha\, b^\beta} = \frac{\Gamma(\alpha + \beta)}{\Gamma(\alpha)
\, \Gamma(\beta)} \int_0^1 {\rm d}x \, \frac{x^{\alpha-1} (1-x)^{\beta-1}}{
[ax + b(1-x)]^{\alpha+\beta}} 
\end{equation}
and shifting ${\bf q} \rightarrow {\bf q} + {\bf k} x$, into
\begin{equation}
{\cal F}_{CT} = - \frac{g^2}{8} \int_0^1 {\rm d}x \, \frac{1}{\sqrt{1-x}}
\int \frac{{\rm d}^3{\bf q}}{(2 \pi)^3} \,  \frac{1}{[{\bf q}^2 + {\bf k}^2
x(1-x)]^{3/2}} \,\, .
\end{equation}
Now compute the last integral in $d = 3 - \epsilon$ dimensions.
Note that this implies $g^2 \rightarrow g^2 \tilde{\Lambda}^{3-d}$, where
$\tilde{\Lambda}$ has dimensions of energy.
The standard formula \cite{thooft}
\begin{equation}
\int_0^\infty {\rm d}x\, x^{\beta} \frac{1}{[x^2 + M^2]^{\alpha}}
= \frac{1}{2}\, \frac{\Gamma\left(\frac{\beta+1}{2}\right) \,
\Gamma\left(\alpha-\frac{\beta+1}{2}\right)}{\Gamma(\alpha) \,
[M^2]^{\alpha-(\beta+1)/2}} 
\end{equation}
leads to
\begin{equation} \label{CT2}
{\cal F}_{CT} = - \frac{g^2}{16 \pi^2} \left[\frac{2}{\epsilon}
- \ln \frac{{\bf k}^2}{\Lambda^2e^2} \right] \,\, ,
\end{equation}
where $\Lambda e$ is the renormalization scale,
$\Lambda \equiv \tilde{\Lambda}\, \sqrt{\pi} \, e^{1+\gamma/2}$,
$\gamma$ being the Euler--Mascheroni constant. 
As usual, the $1/\epsilon$ term in 
(\ref{CT2}) is discarded.


\begin{thebibliography}{99}

\bibitem{fetterwalecka}
A.L.\ Fetter and J.D.\ Walecka, {\it Quantum Theory of Many-Particle Systems}
(McGraw--Hill, New York, 1971); A.A. Abrikosov, L.P.\ Gorkov, and I.E.\ 
Dzyaloshinski, {\it Methods of Quantum Field Theory in Statistical 
Physics} (Dover, New York, 1963).

\bibitem{barrois}
B.C.\ Barrois, Nucl.\ Phys.\ {\bf B129}, 390 (1977).

\bibitem{bailinlove}
D.\ Bailin and A.\ Love, Phys.\ Rep.\ {\bf 107}, 325 (1984).

\bibitem{others}
J.F.\ Donoghue and K.S.\ Sateesh, Phys.\ Rev.\ D {\bf 38}, 360 (1988);
M.\ Iwasaki and T.\ Iwado, Phys.\ Lett.\ {\bf B350}, 163 (1995).

\bibitem{kond}
L.A.\ Kondratyuk, M.M.\ Giannini, and M.I.\ Krivoruchenko,
Phys.\ Lett.\ {\bf B269}, 139 (1991);
L.A.\ Kondratyuk and M.I.\ Krivoruchenko,
Z.\ Phys.\ A {\bf 344}, 99 (1992);
T.\ Sch\"afer, Phys.\ Rev.\ D {\bf 57}, 3950 (1998).

\bibitem{arw}
M.\ Alford, K.\ Rajagopal, and F.\ Wilczek, Phys.\ Lett.\ {\bf B422},
247 (1998).

\bibitem{rssv}
R.\ Rapp, T.\ Sch\"afer, E.V.\ Shuryak, and M.\ Velkovsky,
Phys.\ Rev.\ Lett.\ {\bf 81}, 53 (1998);
T.\ Sch\"afer, Nucl.\ Phys.\ {\bf A642}, 45 (1998).

\bibitem{nambu}
Y.\ Nambu and G.\ Jona-Lasinio, 
Phys.\ Rev.\ {\bf 122}, 345 (1961).

\bibitem{arw2}
M.\ Alford, K.\ Rajagopal, and F.\ Wilczek, Nucl.\ Phys.\
{\bf B537}, 443 (1999).

\bibitem{paterson}
A.J.\ Paterson, Nucl.\ Phys.\ {\bf B190 [FS3]}, 188 (1981);
R.D.\ Pisarski and D.L.\ Stein, J.\ Phys.\ {\bf A14}, 3341 (1981).

\bibitem{pr}
R.D.\ Pisarski and D.H.\ Rischke, nucl-th/9811104.

\bibitem{sw2}
T.\ Sch\"afer and F.\ Wilczek, Phys.\ Rev.\ Lett.\
{\bf 82}, 3956 (1999).

\bibitem{br}
J.\ Berges and K.\ Rajagopal, Nucl.\ Phys.\ {\bf B538}, 215 (1999).

\bibitem{lr}
K.\ Langfeld and M.\ Rho, hep-ph/9811227.

\bibitem{ehs}
N.\ Evans, S.D.H.\ Hsu, and M.\ Schwetz, Nucl.\ Phys.\ {\bf B551}, 
275 (1999); Phys.\ Lett.\ {\bf B449}, 281 (1999).

\bibitem{sw1}
T.\ Sch\"afer and F.\ Wilczek, Phys.\ Lett.\ {\bf B450}, 325 (1999).

\bibitem{son}
D.T.\ Son, Phys.\ Rev.\ D {\bf 59}, 094019 (1999).

\bibitem{QCDrdpdhr} 
R.D.\ Pisarski and D.H.\ Rischke, in preparation.

\bibitem{LeBellac}
M.\ LeBellac, {\it Thermal Field Theory}, (Cambridge University Press,
Cambridge, 1996).

\bibitem{khodel}
V.A.\ Khodel, V.V.\ Khodel, and J.W.\ Clark, Nucl.\ Phys.\ {\bf A598}, 
390 (1996).

\bibitem{dhrwg}
See, for example:
D.H.\ Rischke and W.\ Greiner, Int.\ J.\ Mod.\ Phys.\ {\bf E3}, 1157 (1994).

\bibitem{gross}
F.\ Gross, {\it Relativistic Quantum Mechanics and Field Theory} (Wiley, 
New York, 1993).

\bibitem{thooft}
G.\ t'Hooft and M.\ Veltman, Nucl.\ Phys.\ {\bf B44}, 189 (1972).

\end{thebibliography}
\end{document}